\documentclass[a4paper,12pt,final]{article}
\usepackage[T1]{fontenc}
\usepackage{fullpage}
\usepackage{color}
\usepackage{bm}
\usepackage{amssymb,amsmath}  
\usepackage{graphicx}
\usepackage{mathrsfs} %needed for \mathscr
\usepackage[space]{grffile}
\usepackage[notcite,notref]{showkeys}
\usepackage{color}
\usepackage{fancybox}
\usepackage{paralist} 
\usepackage{mdframed}
\usepackage{float}
\newcommand\mathds[1]{\bm{\mathsf{#1}}}

\newcommand\UU{\mathbf U_{\rm d}}
\newcommand\dotUU{\dot{\mathbf U}_{\rm d}}
\newcommand\CC{\mathbf S_{\rm d}}

\newcommand\blockmatrix[4]{
\left(
\begin{array}{c | c}
#1 & #2\\
\hline
#3 & #4
\end{array}
\right)
}

\title{\color{black}On \color{black}morphoelastic rods}

\author{A. Tiero\footnote{Dipartimento di Ingegneria Civile e Ingegneria Informatica, Universit\`a di Roma Tor Vergata. Via Politecnico 1, 00133 Roma. Italy. Email: \texttt{tiero@uniroma2.it}, \texttt{tomassetti@ing.uniroma2.it}. G.T. acknowledges financial support from the Italian INdAM-GNFM through Project \emph{``PROGETTO GIOVANI 2013: Modellazione matematica dei fenomeni di morphing''}.
}\quad and G. Tomassetti${}^*$}

\date{}
\begin{document}
\maketitle
%\begin{center}
%Dedicated to Prof. Antonio DiCarlo in recognition of his academic activity.
%\end{center}
\begin{abstract}
\color{black}Morphoelastic rods are thin bodies which \color{black}can grow and can change \color{black}their \color{black}intrinsic curvature and torsion. \color{black}We deduce a system of equations ruling accretion and remodeling in a morphoelastic rod \color{black}by combining balance laws involving non-standard forces with constitutive prescriptions filtered by a dissipation principle that takes into account both standard and non-standard working. \color{black}We find that, as in the theory of three-dimentional bulk growth proposed in [A. DiCarlo and S. Quiligotti, Mech. Res. Commun. 29 (2002) 449-456], it is possible to identify a universal coupling mechanism between stress and growth, conveyed by an Eshelbian driving force.\color{black} 

\end{abstract}

 \section{Introduction}
Unlike common engineering materials, living matter can adapt to environmental changes by growing and by actively modifying its structure. When trying to accommodate these features in the infrastructure of continuum mechanics, a key issue is to distinguish growth from strain. Beginning with \cite{RodriHM1994JB}, this issue has been addressed through the multiplicative decomposition
\begin{equation}\label{eq:66}
\color{black}\nabla{\boldsymbol\chi}\color{black}=\bm{\mathbf  F}\bm{\mathbf  G}
\end{equation}
of the \emph{deformation gradient} $\nabla\color{black}\boldsymbol\chi$ into a part $\mathbf F$ accounting for mechanically-induced strain, and a part $\mathbf G$ accounting for growth.  The multiplicative decomposition \eqref{eq:66} plays a central role in several mechanical theories, such as finite-deformation thermoplasticity and elastoplasticity \cite{Lubar2004AMR}. In all these theories, $\mathbf G$ 
%%%%%%%%%%%%%%%%%%%%
%is interpreted as 
yields
%%%%%%%%%%%%%%%%%%%%
the local \emph{zero-stress state}, and its value at a given point of the reference configuration indicates how the neighborhood of that point would deform if isolated from the rest of the body.

The additional degrees of freedom brought in by the \emph{growth tensor} $\mathbf G$ demand additional evolution laws whose choice and interpretation set a formidable challenge. In the format set forth in \cite{DiCarQ2002MRC} the laws governing the evolution of $\mathbf G$ are obtained by combining suitable constitutive prescriptions with the balance equation
\begin{align}\label{eq:19}
  \mathbf C=\bm{\mathbf   B},
\end{align}
which involves an \emph{inner remodeling couple} $\mathbf C$ and an \emph{outer remodeling couple} $\bm{\mathbf   B}$. Remodeling couples are work conjugates of the \emph{accretion velocity}:
\begin{align}
\mathring{\bm{\mathbf   G}}=\dot{\bm{\mathbf   G}}{\bm{\mathbf   G}}^{-1}
\end{align}
(the superimposed dot denotes partial derivative with respect to time), and cooperate with \color{black}standard stress and \color{black}standard forces to the total power expenditure. The balance statement \eqref{eq:19}, obtained from the principle of virtual powers along the guidelines set forth by Germain \color{black}\cite{Germa1973JM,Germa1973SJAM}\color{black}, is accompanied by a carefully-tailored version of the dissipation inequality. \color{black}If the stress respose of the material is elastic, the dissipation inequality is satisfied during whatever evolution process if and only if the self-remodeling couple has the form:
\begin{align}
  \mathbf C=\bm{\mathbf  E}+\mathbf C_+
\end{align}
where the dissipative response  ${\mathbf C}_+$ expends non-negative power, and
\begin{align}\label{eq:74}
  \bm{\mathbf   E}=\widehat\varphi(\bm{\mathbf   F})\,\bm{\mathbf   I}-\bm{\mathbf   F}^{\rm T}\mathbf S,\qquad 
\end{align}
with $\widehat\varphi(\mathbf F)$ the \emph{free energy per unit relaxed volume} and $\mathbf S=\partial\check\varphi(\mathbf F)$ the \emph{energetic stress}.

\color{black}Remarkably, the expression of the \color{black}\emph{energetic self-remodeling couple} \color{black}$\mathbf E$ coincides, \emph{mutatis mutandis}, with that of the \emph{Eshelby stress}, a tensorial quantity involved in Eshelby's notion of a force acting on a defect in an elastic body \cite{Eshel2006Continuum}. Eshelby's original derivation was based on variational arguments, and it took quite a while to reckon \eqref{eq:74}  as a constitutive statement for a non-standard dynamical decriptor  \cite{Gurti2000a,Maugi1993,Podio2002MRC}. Yet, several researchers still contend the mechanical relevance of the Eshelby stress, regarding it as a derived object. As a matter of fact, it is not necessary to invoke the theory of configurational forces to argue that the Eshelby tensor has a mechanical significance of its own: genuine variational arguments can be adduced to show that if material structure is allowed to change then standard and configurational balances cannot be deduced from each other \cite{Podio2001IFB}.

The literature on three-dimensional continuous bodies which can change their relaxed shape by growth and remodeling is extremely developed, as evident from recent surveys on the subject \cite{AmbroAACDGHHKKOTG2011JMPS,Humph2003PotRSoLSAMPaES,MenzeK2012MRC}. This is not the case for \emph{morphoelastic structures} \color{black}that is to say, thin bodies, such as rod and shells, which can change their relaxed shape \color{black}\cite{GoldsG2006PRE,McMahGT2011MaMoS,OReiT2011IJSS,MoultLG2013JMPS}\color{black}. This, despite the fact that models of thin structures, being more accessible to analytical investigation because of their simplicity, seem to be better suited to explain certain qualitative feature of several biological systems \cite{GorieB2005PRL,GorieT1998PRL,GuillDF2012MCM}. 

When considering a rod, one may ask: 1) what decomposition should replace \eqref{eq:66}; what balance statement should take the place of \eqref{eq:19}; 3) what mechanical construct would substitute the Eshelby stress. When looking for an answer to these questions, one must recall that a rod is a one-dimensional oriented continuum; as such, its configuration consists of a pair $\mathsf r=(\mathbf R,\mathbf r)$ of a vector field $\mathbf r$ and a rotation field $\mathbf R$ defined in one-dimensional domain (the axis of the rod in the reference configuration). The vector $\mathbf r(x)$ delivers the placement of the centroid of the typical section $x$, whereas the rotation $\mathbf R(x)$ yields the orientation of the same section (see Fig. 1 and Eq. \eqref{eq:1} below). For an \emph{unshearable rod}, the case we consider in this paper, $\mathbf r$ and $\mathbf R$ are related by 
\begin{align}\label{eq:75}
 \frac{\partial_x\mathbf r(x)}{|\partial_x\mathbf r(x)|}
=\mathbf R(x)\mathbf a,
\end{align}
where $\mathbf a$ is a constant unit vector orthogonal to the section in the reference configuration (see Fig. 2 below). 
%%%%%%%%%%%%%%%55
%We are then 
Then, one is  
%%%%%%%%%%%%%
led to mimick \eqref{eq:66} by decomposing the \emph{visible stretch} $|\partial_x\mathbf r(x)|$ into a \emph{mechanical stretch} $\nu>0$ and\color{black}, using the terminology of \cite{MoultLG2013JMPS}, \color{black}a \emph{growth stretch} $\lambda>0$:
\begin{align}\label{eq:76}
  |\partial_x\mathbf r(x)|=\nu\lambda.
\end{align}
In order to provide the theory with a non-trivial range of application, we cannot content ourselves with \eqref{eq:76}. In actual biological systems, not only length, but also spontaneous curvature and torsion evolve with time. For instance, the change of spontaneous curvature is the mechanism at the basis of gravitropism \cite{BastiBMD2013PNASU,FournBC1994B,MouliCL2006AJB}, namely, the ability of certain plant organs to attain and keep their vertical posture. In  the so-called \emph{geometrically-exact rod model} \cite{Simo1985CMAM}, curvature and torsion are accounted for through the skew-symmetric tensor 
\begin{align}\label{eq:81}
  \mathbf U=\mathbf R^T\mathbf R',
\end{align}
where the prime sign denotes differentiation along the axis in the relaxed configuration:\footnote{Compare with (8) in \cite{DiCar2005Surface}.}
\begin{align}\label{eq:77}
  \mathbf R'=\lambda^{-1}\partial_x\mathbf R.
\end{align}
Observe that, for a standard \color{black}elastic \color{black}rod, $\lambda=1$ so that $\mathbf R'=\partial_x\mathbf R$ becomes the conventional derivative along the axis in the reference configuration (which in this case coincides with the relaxed configuration). In order to allow the curvature of the relaxed configuration to change with time, we decompose the strain measure $\mathbf U$ into a mechanical part $\mathbf U_{\rm m}$ and a distortion part $\mathbf U_{\rm d}$:
\begin{align}\label{eq:80}
  \mathbf U=\mathbf U_{\rm m}+\mathbf U_{\rm d},
\end{align}
and we assume that the free energy per unit relaxed length depends only on the mechanical part of $\mathbf U$:
\begin{align}
  \psi=\widehat\psi(\mathbf U_{\rm m}).
\end{align}
When comparing \eqref{eq:80} with the decompositions \eqref{eq:66} and \eqref{eq:76}, one cannot but notice the addition operation in place of composition. The reason for this discrepancy is that the strain measure $\mathbf U$ should be thought as an element of a linear space, rather than of a group. Indeed, if we think of $\mathbf R$ as an element of a differentiable manifold (namely, the \emph{Lie group} $\textrm{SO}(3)$), then by a glance at \eqref{eq:81} see that $\mathbf U$ is an element of the tangent space at the identity $\mathbf I$ (precisely, the \emph{Lie algebra} $\mathfrak{so}(3)$). 

\color{black}We remark that decomposition such as \eqref{eq:76} and \eqref{eq:80} are standard in structural theories that incorporate anelastic effects. However, what distinguishes growth stretch from thermal or plastic stretch is its use as renomalization factor to define energy and work densities (see for instance \eqref{eq:3}, \eqref{eq:4}, and \eqref{eq:5} in Section 2.2 of this paper).\color{black} 

With \eqref{eq:76} and \eqref{eq:80} at our disposal, we recognize the scalar field $\lambda$ and the tensor field $\mathbf U_{\rm d}$ as the additional degrees of freedom brought in by growth and remodeling. Accordingly, we introduce an \emph{internal accretion force} $S$, work conjugate of the \emph{axial accretion velocity}
\begin{align}
  \mathring\lambda=\frac{\dot\lambda}\lambda,
\end{align}
 and a \emph{remodeling moment} $\mathbf S_{\rm d}$, work conjugate to the \emph{remodeling velocity} $\dot{\mathbf U}_{\rm d}$. 
%%%%%%%%
%We 
In order to make our treatment as simple as possible, we
%%%%%%%%
neglect elastic stretch by setting 
%%%%%%%%
%$\nu=0$,
\begin{align}
  \nu=1,
\end{align}
%%%%%%%%
so that $\mathbf U$ is the only visible strain. The evolution of $\mathbf U$ is accompanied by internal power expended by a skew symmetric \emph{bending-torsion moment} $\mathbf S$ (see Fig. 3 below). 
%%%%%%%%
Thus the internal power per unit content is:
\begin{align}
  w_{\rm int}=\mathbf S\cdot\dot{\mathbf U}+\mathbf S_{\rm d}\cdot\dot{\mathbf U}_{\rm d}+S\mathring\lambda.
\end{align}
%%%%%%
Virtual variations of the growth stretch $\lambda$ and of the distortion field $\mathbf U_{\rm d}$ generate the balance laws (see \eqref{eq:13} and \eqref{eq:84} below):
\begin{subequations}\label{eq:82}
\begin{align}
 & S-\mathbf S\cdot\mathbf U-\mathbf s\cdot\mathbf u=B,\\
 &\mathbf S_{\rm d}=\mathbf B_{\rm d},
\end{align}
\end{subequations}
where $\mathbf s$ is the (reactive) work-conjugate of the vectorial strain measure $\mathbf u$  (see Fig. 2 and Eq. \eqref{eq:8} below) associated to shear and extension, which in the present case coincides with the unit vector $\mathbf a$ (see \eqref{eq:9} below). In our case, the energetic part the internal accretion force coincides with the free energy per unit relaxed length: $S^{\rm ener}=\widehat\psi(\mathbf U_{\rm m})$. Thus, the balance equation (\ref{eq:82}a) takes the form
\begin{align}
  E+S^{\rm diss}=B,
\end{align}
where $S^{\rm diss}$, the dissipative part of the internal accretion force, enters a reduced dissipation inequality (see \eqref{eq:18} below), and where
\begin{align}\label{eq:89}
  E=\widehat\psi(\mathbf U_{\rm m})-\mathbf S\cdot\mathbf U-\mathbf s\cdot\mathbf u
\end{align}
is the \emph{Eshelbian accretion force}.

The theory we propose should be suitable to model biological system whose geometrical features resemble those of a rod, such as for example plant organs. In particular, the external forces $B$ and ${\mathbf B_{\rm d}}$ may be used to account for physical and chemical stimuli from the environment, which drive growth and remodeling. Their choice depends on the particular application one is being after, and is beyond the scope of the present paper. Rather, our aim is to emphasize that a state of stress alone may induce growth and remodeling, even in the absence of chemical actions. In order to illustrate this point, we provide two examples in Section 3. In Example 1, growth takes place at the expense of external work performed by an applied load. In Example 2, growth is induced by relaxation of elastic energy. In both cases, an important role is played by the \emph{cost of axial accretion}:
\begin{align}
  \psi_0=\widehat\psi(\mathbf 0),
\end{align}
namely, the energetic cost associated to a unit increase of the growth stretch $\lambda$. \color{black}In particular, it is seen that if this cost is positive, then the structure has a tendency to shrink, unless other material forces prevent it from doing so.\color{black}

\color{black}What distinguishes our approach from other treatments of morphoelastic rods is that in our model the laws governing the evolution of the spontaneous curvature and growth stretch are a consequence of a constitutively-augmented non-standard force balances. In particular, the Eshelbian accretion force defined in \eqref{eq:89} provides a possible mechanism for the interaction between growth and stress descriptors. As a matter of fact the notion of material force in structural theories is not new \cite{Braun2005Structural,KienzH1986AJAM,KienzH1986IJSS,ORei2007JE}. In the small-strain setting, the configurational balance equation \eqref{eq:13} has already been deduced in \cite{Tomas2011AAM} by extending previous work in \cite{PedePT2006NCB,Podio2005Peeling}. Very recently, theoretical insight and experimental evidence has shown the relevance of material forces in rods \cite{BigonCBM2013MM}.\color{black}

\section{General theory}
\subsection{Visible kinematics}
We identify the reference configuration of the rod with the cylinder $\mathcal R=(0,1)\times\mathcal S$ where $\mathcal S$ is a plane region. Thus, we think the typical material point of $\mathcal R$ as a pair $(x,p)$, with  $x\in (0,1)$ and $p\in\mathcal S$. As in standard rod theory, we assume that the place $f$ occupied by this point at time $t$ has the following representation:
\begin{equation}
  \label{eq:1}
  f(x,p,t)=q_0(x,t)+\mathbf R(x,t)(p-p_0),\qquad q_0(x,t)=p_0+\mathbf r(x,t),
\end{equation}
where $\mathbf r(x,t)$ is a vector and $\mathbf R(x,t)$ is a rotation tensor.
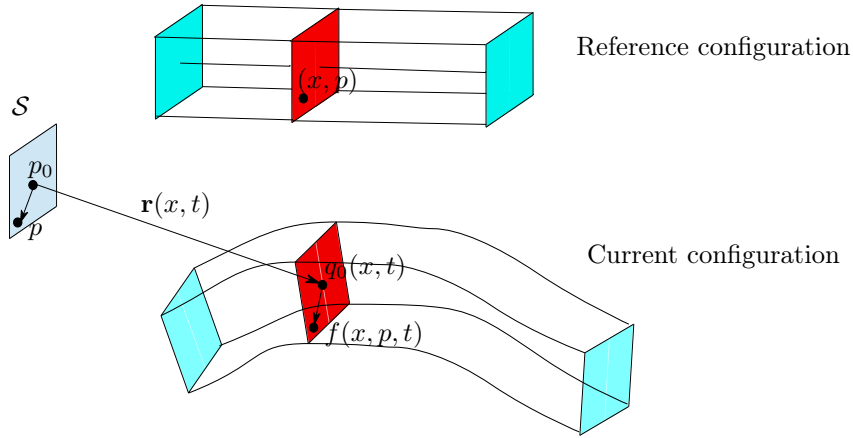
\begin{figure}[H]
\begin{center}
\footnotesize
\def\svgwidth{0.6\linewidth}
%% Creator: Inkscape inkscape 0.48.4, www.inkscape.org
%% Accompanies image file 'figure-motion.pdf' (pdf, eps, ps)
\begingroup%
  \makeatletter%
  \providecommand\color[2][]{%
    \errmessage{(Inkscape) Color is used for the text in Inkscape, but the package 'color.sty' is not loaded}%
    \renewcommand\color[2][]{}%
  }%
  \providecommand\transparent[1]{%
    \errmessage{(Inkscape) Transparency is used (non-zero) for the text in Inkscape, but the package 'transparent.sty' is not loaded}%
    \renewcommand\transparent[1]{}%
  }%
  \providecommand\rotatebox[2]{#2}%
  \ifx\svgwidth\undefined%
    \setlength{\unitlength}{793.44561709bp}%
    \ifx\svgscale\undefined%
      \relax%
    \else%
      \setlength{\unitlength}{\unitlength * \real{\svgscale}}%
    \fi%
  \else%
    \setlength{\unitlength}{\svgwidth}%
  \fi%
  \global\let\svgwidth\undefined%
  \global\let\svgscale\undefined%
  \makeatother%
  \begin{picture}(1,0.59945468)%
    \put(0,0){\includegraphics[width=\unitlength]{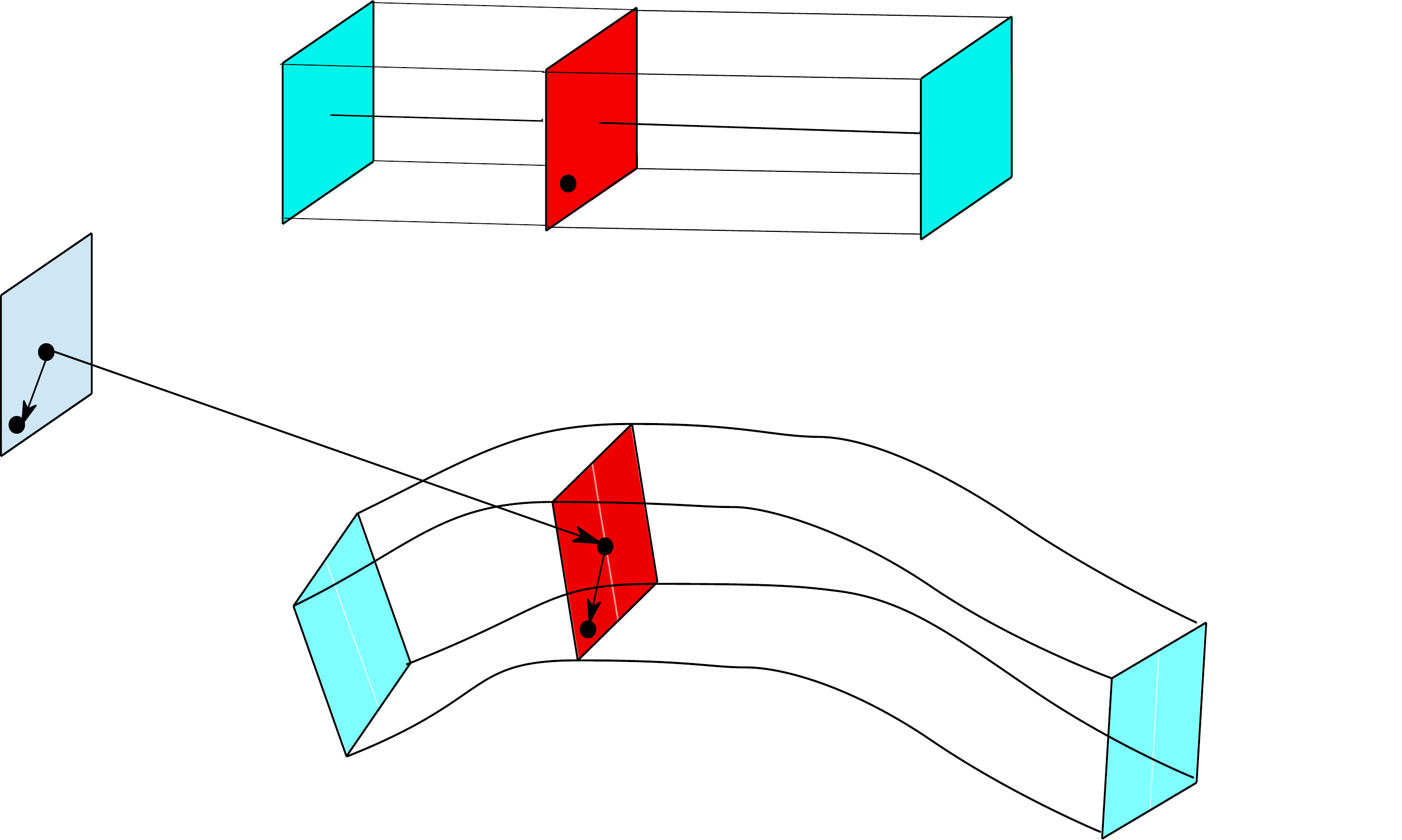}}%
    \put(0.02684866,0.36439005){\color[rgb]{0,0,0}\makebox(0,0)[lb]{\smash{$p_0$}}}%
    \put(0.43471353,0.22459974){\color[rgb]{0,0,0}\makebox(0,0)[lb]{\smash{$q_0(x,t)$}}}%
    \put(0.43471356,0.13224298){\color[rgb]{0,0,0}\makebox(0,0)[lb]{\smash{$f(x,p,t)$}}}%
    \put(0.02716333,0.2782536){\color[rgb]{0,0,0}\makebox(0,0)[lb]{\smash{$p$}}}%
    \put(0.18171288,0.31118092){\color[rgb]{0,0,0}\makebox(0,0)[lb]{\smash{$\mathbf r(x,t)$}}}%
    \put(0.3964968,0.48256148){\color[rgb]{0,0,0}\makebox(0,0)[lb]{\smash{$(x,p)$}}}%
    \put(0.00380707,0.44528151){\color[rgb]{0,0,0}\makebox(0,0)[lb]{\smash{$\mathcal S$}}}%
    \put(0.78184714,0.52873982){\color[rgb]{0,0,0}\makebox(0,0)[lb]{\smash{Reference configuration}}}%
    \put(0.79617837,0.24052325){\color[rgb]{0,0,0}\makebox(0,0)[lb]{\smash{Current configuration}}}%
  \end{picture}%
\endgroup%
\footnotesize\caption{Placement of the typical section.}
\end{center}  
\end{figure}
\noindent Accordingly, the velocity of the same point can be written as $$\dot f(x,p,t)=\mathbf w(x,t)+\mathbf W(x,t)(f(x,p,t)-q_0(x,t)),$$ where
\begin{align}\label{eq:2}
\mathbf W=\dot{\mathbf R}\mathbf R^T\qquad\textrm{and}\qquad \mathbf w=\dot{\mathbf r}-\mathbf W\mathbf r
\end{align}
are, respectively, the \emph{spin} and the \emph{linear velocity} resolved with respect to
$p_o$. Our motivation for resolving velocities with respect to the \emph{fixed pole}
$p_o$ instead of the \emph{moving pole} $q_o(x,t)$ will be apparent when discussing balance
equations.

\subsection{Kinematics of growth and remodeling}

Three traits distinguish our developments from standard rod theory. The first is that, when computing energies and power expenditures as integrals of certain densities, we do not perform integration with respect to the Lebesgue measure ${\rm d}x$. Instead, we integrate with respect to the time-dependent \emph{content measure}:
\begin{equation}
  \label{eq:3}
  {\rm d}s=\lambda(x,t){\rm d}x,
\end{equation}
whose density $\lambda$, the \emph{specific content}, is a positive scalar field that is prescribed only at the beginning of the motion, and evolves according to certain laws that we shall specify later in this paper.  To give two examples of this practice: 1) we write the free energy stored in a \emph{part} $(a,b)\subset(0,1)$ as 
\begin{equation}
  \label{eq:4}
\Psi[(a,b)]:=\int_a^b\psi {\rm d}s,  
\end{equation}
and we refer to $\psi$ as the \emph{free energy per unit content}; 2) when prescribing the power expenditure of body forces on $(a,b)$, we write
\begin{equation}
  \label{eq:5}
  \mathcal W_{\rm ext,b}[(a,b)]:=\int_a^b \big(\mathbf B\cdot\mathbf W+\mathbf b\cdot\mathbf w\big){\rm d}s.
\end{equation}
We refer to $\mathbf B$ and $\mathbf b$, as respectively, the \emph{body couple resolved with respect to $p_o$} and the \emph{body force}, measured per per unit content. Here the dot product between skew tensors is defined as $\mathbf B\cdot\mathbf W=\color{black}\frac 12 \mathbf B:\mathbf W=\color{black}\frac 1 2{\rm tr}(\mathbf B^T\mathbf W)$, with ${\rm tr(\cdot)}$ the \emph{trace operator}. \footnote{\color{black}A consequence of adopting this convention is that the \color{black}dot product between two skew tensors coincides with the scalar product between the corresponding axial vectors. For instance, on denoting by $\bm b$ and $\bm w$ the axial vectors 
of, respectively, $\mathbf B$ and $\mathbf W$, we have $\mathbf B\cdot\mathbf W=
\bm b\cdot\bm w$.}

\color{black}
\smallskip

A second distinguishing trait of our theory is that, when performing derivatives holding $t$ fixed, we make systematic use of the differential operator:
\begin{equation}
  \label{eq:6}
  \varphi'=\lambda^{-1}\partial_x\varphi.
\end{equation}
The significance of the differential operator $(\cdot)'$ is best understood by considering the parametrization  
\begin{equation}\label{eq:7}
x\mapsto s=s(x,t)=\int_0^x\lambda(y,t){\rm d}y.
\end{equation} 
The field $s(x,t)$ is \emph{the total content of the part $(0,x)$ at time $t$}. 
It is readily seen that $\varphi'$ is the derivative of $\varphi$ with respect to
 $s$ holding $t$ fixed. \footnote{Since $\lambda$ is strictly positive,this function has an inverse $s\mapsto \widetilde x(s,t)$. Hence, the description $\varphi(x,t)$ of any time dependent field $\varphi$  may be provided in terms of $s$ through the function $\widetilde\varphi(s,t):=\varphi(\widetilde x(s,t),t)$, and $\varphi'(x,t)=\partial_s\widetilde\varphi(s(x,t),t)$.}
%and, with slight abuse of notation, we may write $\varphi'=\partial_s\varphi$. 
%In the same fashion, 
Likewise,  
when computing certain derivatives holding $x$ fixed to obtain rate-like quantities, we make use of the operator:
\begin{equation}
  \label{eq:68}
  \mathring\varphi=\lambda^{-1}\partial_t\varphi.
\end{equation}

There are several reasons for introducing the operator $(\cdot)'$. One, purely technical, is that when performing integration with respect to the parameter $s$, integration-by-parts formulas hold true if we differentiate with respect to the same parameter. The other, in our opinion more fundamental, is that it seems to us appropriate to compute derivatives with respect to the content measure when introducing the strain measures:
\begin{equation}
  \label{eq:8}
  \mathbf U=\mathbf R^T\mathbf R',\qquad \mathbf u=\mathbf R^T\mathbf r'.
\end{equation}
The skew-symmetric valued tensor field $\mathbf U(x,t)$ accounts for torsion and bending, the vector field $\mathbf u(x,t)$ 
%accounts 
\color{black}for extension and shear. 

In order to 
%render 
make 
our developments
as simple as possible, 
we 
rule out 
extension and 
%bending 
shear
by 
%imposing 
enforcing the \emph{internal constraint}:
\begin{equation}
  \label{eq:9}
  \mathbf u(x,t)=\mathbf a,
\end{equation}
where $\mathbf a$ is a fixed unit vector orthogonal to $\mathcal S$. In view of \eqref{eq:6}, the constraint \eqref{eq:9} entails $\partial_x\mathbf r(x,t)=\lambda(x,t)\mathbf R(x,t)\mathbf a$. Thus, $\lambda=|\partial_x\mathbf r|$ and, by \eqref{eq:7}, the total content of the part $(0,x)$ is the length of the curve $(0,x)\ni y\mapsto q_0+\mathbf r(y,t)$; moreover, the tangent to the axis at $x$ is perpendicular to the section in the current configuration. 

The third important feature of our theory is that a skew-symmetric \emph{distortion field} $\UU(x,t)$ joins $\mathbf U$ and $\lambda$ in the list $\upsilon=(\mathbf U,\UU,\lambda)$ of \emph{state variables}. The distortion field delivers the intrinsic curvature and the intrinsic torsion of the rod. As pointed out in the Introduction, the pair $(\lambda,\UU)$ plays, in the present context, the same role as the growth tensor $\bm{\mathsf G}$ in the decomposition \eqref{eq:66}.

\begin{figure}[H]\label{fig:12}
\begin{center}
\footnotesize
\def\svgwidth{0.6\linewidth}
%% Creator: Inkscape inkscape 0.48.4, www.inkscape.org
%% Accompanies image file 'figure-strains.pdf' (pdf, eps, ps)
\begingroup%
  \makeatletter%
  \providecommand\color[2][]{%
    \errmessage{(Inkscape) Color is used for the text in Inkscape, but the package 'color.sty' is not loaded}%
    \renewcommand\color[2][]{}%
  }%
  \providecommand\transparent[1]{%
    \errmessage{(Inkscape) Transparency is used (non-zero) for the text in Inkscape, but the package 'transparent.sty' is not loaded}%
    \renewcommand\transparent[1]{}%
  }%
  \providecommand\rotatebox[2]{#2}%
  \ifx\svgwidth\undefined%
    \setlength{\unitlength}{343.75942942bp}%
    \ifx\svgscale\undefined%
      \relax%
    \else%
      \setlength{\unitlength}{\unitlength * \real{\svgscale}}%
    \fi%
  \else%
    \setlength{\unitlength}{\svgwidth}%
  \fi%
  \global\let\svgwidth\undefined%
  \global\let\svgscale\undefined%
  \makeatother%
  \begin{picture}(1,0.87655777)%
    \put(0,0){\includegraphics[width=\unitlength]{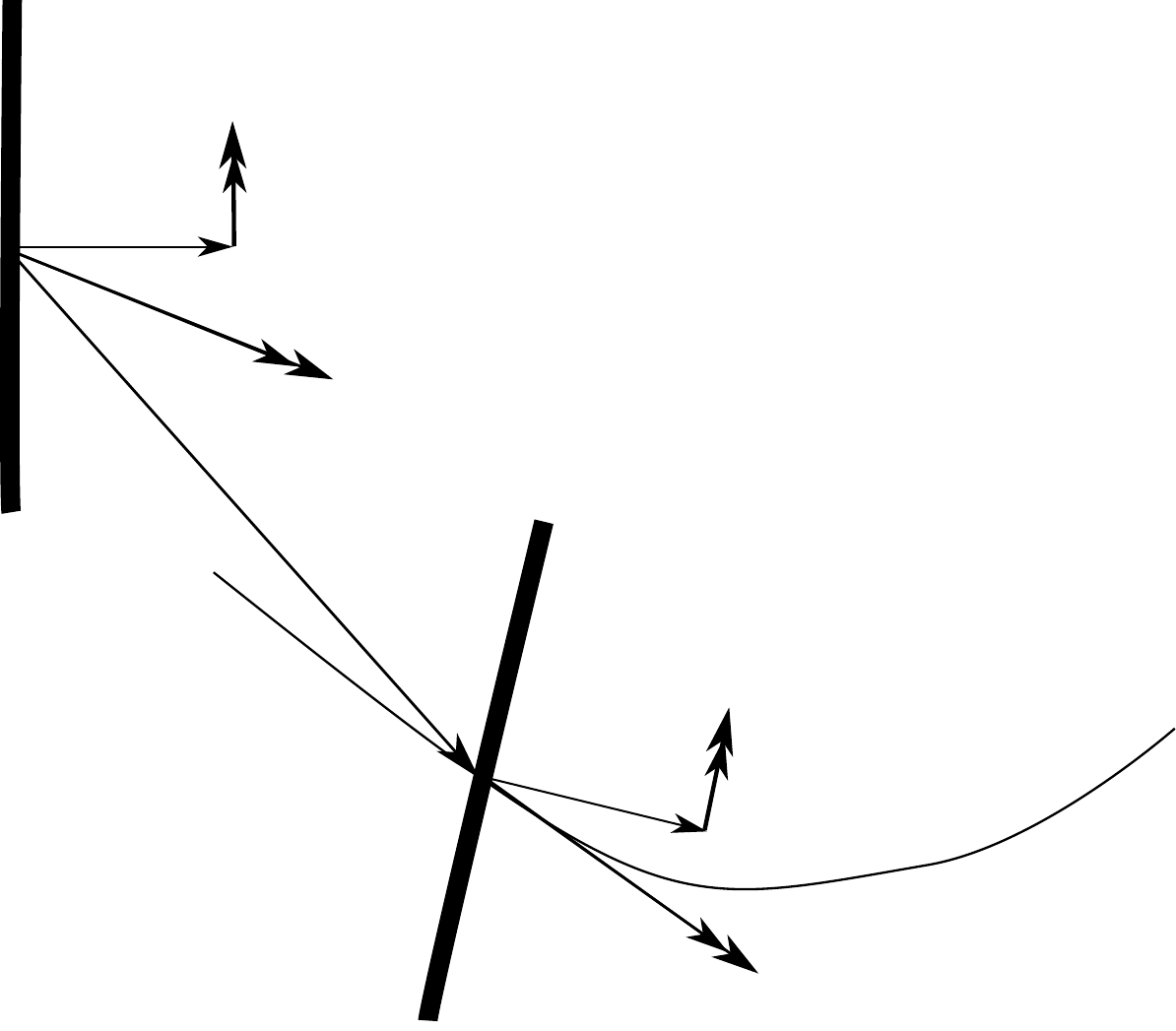}}%
    \put(0.07063664,0.6852111){\color[rgb]{0,0,0}\makebox(0,0)[lb]{\smash{$\mathbf a$}}}%
    \put(0.22278616,0.7127372){\color[rgb]{0,0,0}\makebox(0,0)[lb]{\smash{$\mathbf R^T\mathbf R'\mathbf a$}}}%
    \put(0.09609248,0.51668845){\color[rgb]{0,0,0}\makebox(0,0)[lb]{\smash{$\mathbf R^T\mathbf r'$}}}%
    \put(0.62634789,0.19960873){\color[rgb]{0,0,0}\makebox(0,0)[lb]{\smash{$\mathbf R'\mathbf a$}}}%
    \put(0.48616303,0.05262398){\color[rgb]{0,0,0}\makebox(0,0)[lb]{\smash{$\mathbf r'$}}}%
    \put(0.44533695,0.22169094){\color[rgb]{0,0,0}\makebox(0,0)[lb]{\smash{$\mathbf R\mathbf a$}}}%
    \put(0.20220427,0.4342344){\color[rgb]{0,0,0}\makebox(0,0)[lb]{\smash{$\mathbf r$}}}%
  \end{picture}%
\endgroup%
\caption{\footnotesize The geometrical interpretation of the strain measures $\mathbf U=\mathbf R^T\mathbf R'$ and $\mathbf u=\mathbf R^T\mathbf r'$. We use double arrows to denote vectors obtained by differentiation (as well as their pull backs). The component of $\mathbf u$ along $\mathbf a$ measures axial stretch; the orthogonal component measures shear. If the rod is inextensible, then $\mathbf r'$ is a unit vector. If, in addition, the rod is unshearable, then $\mathbf r'=\mathbf R\mathbf a$, and hence $\mathbf u=\mathbf a$.}
\end{center}
\end{figure}
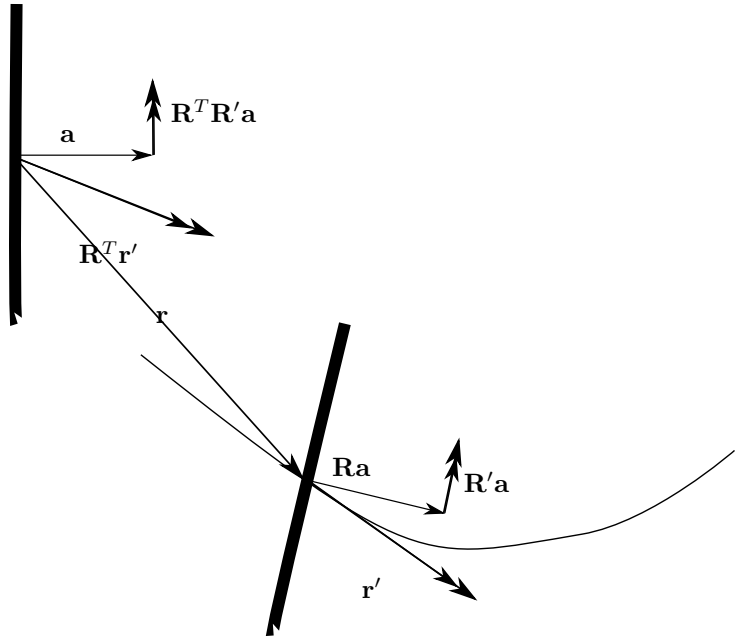
%The kinematical interpretation of \eqref{eq:9} is the following. 
% the function $s$ introduced in  \eqref{eq:7} is the arc-length parameter on the \emph{axis} $x\mapsto q_0+\mathbf r(x,t)$
% In view of \eqref{eq:6}, the constraint \eqref{eq:9} 
% %can be written as 
% \color{black}yields \color{black}$\partial_x\mathbf r(x,t)=\lambda(x,t)\mathbf R(x,t)\mathbf a$. Its geometrical interpretation is the following: 1) the function $s$ introduced in  \eqref{eq:7} is the arc-length parameter on the \emph{axis} $x\mapsto q_0+\mathbf r(x,t)$; 2) the tangent to the axis at $x$ is perpendicular to $\mathcal S_x(t)$. 
\smallskip

%It is important to point out, however, that $\lambda$ accounts for growth, whereas $\UU$ accounts for remodeling. 

\subsection{Internal and external powers}
Adopting the point of view of \cite{DiCarQ2002MRC}, we associate to the evolution of the \emph{state}  $\upsilon=(\mathbf U,\UU,\lambda)$ the \emph{internal power}
\begin{equation}
  \label{eq:111}
\mathcal W_{\rm int}([a,b])=\int_a^b \big(\mathbf S\cdot\dot{\mathbf U}+\CC\cdot\dotUU+S\mathring\lambda\big){\rm d}s
\end{equation}
expended within every part $(a,b)\subset(0,1)$ by a system of \emph{dynamical descriptors}  $\sigma=(\mathbf S,\CC,S)$. We require that internal power expenditure be balanced by an external power, which we assume to be of the form:
\begin{equation}
  \label{eq:11}
  \mathcal W_{\rm ext}[(a,b)]=\int_a^b \big(\mathds b\cdot\mathds w+{{\mathbf B_{\rm d}}}\cdot\dotUU+B\mathring\lambda\big){\rm d}s+\mathds t\cdot\mathds w|_{b}^a,
\end{equation}
where
\begin{align}\label{eq:63}
\mathds w=(\mathbf W,\mathbf w),\quad \mathds b=(\mathbf B,\mathbf b),\quad \mathds t=(\mathbf T,\mathbf t).
\end{align}
Besides power expenditure by external body forces (\emph{cf.} \eqref{eq:5}), the right-hand side of \eqref{eq:11} contains power expenditure by the external forces ${\mathbf B_{\rm d}}$ and $B$ driving remodeling and growth. The field $\mathds t$ conveys the mechanical contact actions exerted on a part $(a,b)$ by its complement $(0,a)\cup (b,0)$ (the rest of the rod). Thus, $\int_a^b\mathbf b{\rm d}s+\mathbf t|_{a}^b$ and $\int_a^b\mathbf B{\rm d}s+\mathbf T|_a^b$ represent, respectively, the resultant force acting on $(a,b)$ and the skew tensor associated to the resultant moment \emph{resolved with respect to the pole $p_o$}.

\begin{figure}[H]\label{fig:10}
\begin{center}
\footnotesize
\def\svgwidth{0.6\linewidth}
%% Creator: Inkscape inkscape 0.48.4, www.inkscape.org
%% Accompanies image file 'figure-stresses.pdf' (pdf, eps, ps)
\begingroup%
  \makeatletter%
  \providecommand\color[2][]{%
    \errmessage{(Inkscape) Color is used for the text in Inkscape, but the package 'color.sty' is not loaded}%
    \renewcommand\color[2][]{}%
  }%
  \providecommand\transparent[1]{%
    \errmessage{(Inkscape) Transparency is used (non-zero) for the text in Inkscape, but the package 'transparent.sty' is not loaded}%
    \renewcommand\transparent[1]{}%
  }%
  \providecommand\rotatebox[2]{#2}%
  \ifx\svgwidth\undefined%
    \setlength{\unitlength}{671.37058716bp}%
    \ifx\svgscale\undefined%
      \relax%
    \else%
      \setlength{\unitlength}{\unitlength * \real{\svgscale}}%
    \fi%
  \else%
    \setlength{\unitlength}{\svgwidth}%
  \fi%
  \global\let\svgwidth\undefined%
  \global\let\svgscale\undefined%
  \makeatother%
  \begin{picture}(1,0.45049941)%
    \put(0,0){\includegraphics[width=\unitlength]{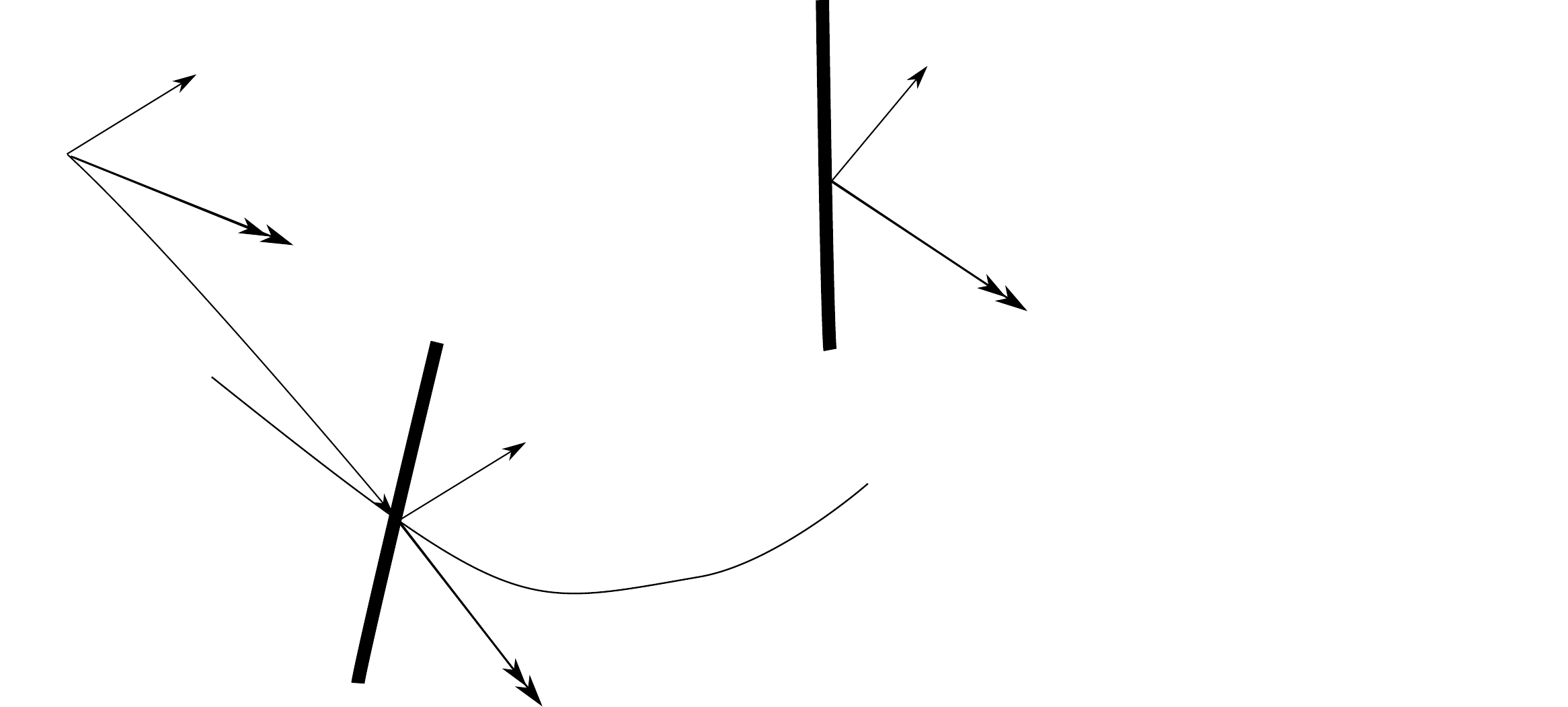}}%
    \put(0.17789799,0.30767342){\color[rgb]{0,0,0}\makebox(0,0)[lb]{\smash{$\mathbf T$}}}%
    \put(0.12928764,0.38228116){\color[rgb]{0,0,0}\makebox(0,0)[lb]{\smash{$\mathbf t$}}}%
    \put(0.34113708,0.02959999){\color[rgb]{0,0,0}\makebox(0,0)[lb]{\smash{$\mathbf M=\mathbf T-\mathbf r\wedge\mathbf t$}}}%
    \put(0.1586668,0.2318199){\color[rgb]{0,0,0}\makebox(0,0)[lb]{\smash{$\mathbf r$}}}%
    \put(0.33922031,0.14876529){\color[rgb]{0,0,0}\makebox(0,0)[lb]{\smash{$\mathbf t$}}}%
    \put(-0.00128585,0.32925032){\color[rgb]{0,0,0}\makebox(0,0)[lb]{\smash{$p_0$}}}%
    \put(0.60403209,0.40996603){\color[rgb]{0,0,0}\makebox(0,0)[lb]{\smash{$\mathbf s=\mathbf R^T\mathbf t$}}}%
    \put(0.63756453,0.27996752){\color[rgb]{0,0,0}\makebox(0,0)[lb]{\smash{$\mathbf S=\mathbf R^T\mathbf M\mathbf R$}}}%
    \put(0.19428768,0.11502368){\color[rgb]{0,0,0}\makebox(0,0)[lb]{\smash{$q_0$}}}%
  \end{picture}%
\endgroup%
\caption{\footnotesize In standard rod theory, the internal forces transmitted by a section are accounted for by their resultant force $\mathbf t$ and their resultant moment $\mathbf M$ with respect to the current position $q_0$ of its centroid. Here we use double arrows to sketch axial vectors of skew-symmetric tensors. The pair $(\mathbf T,\mathbf t)$ represents the same system of internal forces, resolved with respect to the fixed pole $p_0$. Standard transport \emph{formulae} yield $\mathbf T=\mathbf M+\mathbf r\wedge \mathbf t$. The pair $(\mathbf S,\mathbf s)$ represents the same force system, resolved with respect to $q_0$, as registered by an observer moving with the section.}
\end{center}
\end{figure}
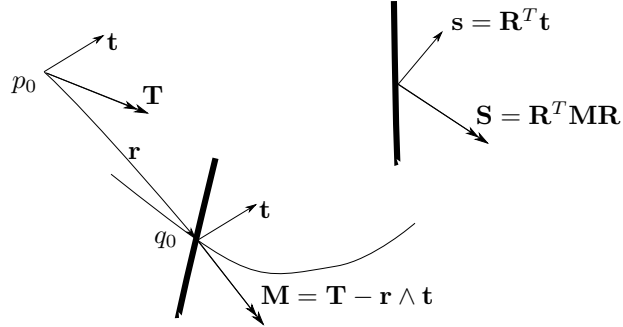
The relation between the various strain and stress measures is best understood by noticing that, when $\mathring\lambda=0$ (no axial growth), the following identities hold (see Appendix 2):
\begin{equation}
\begin{aligned}
  \mathbf S\cdot\dot{\mathbf U}+\mathbf s\cdot\dot{\mathbf u}&=\mathbf T\cdot\mathbf W'+\mathbf t\cdot\mathbf w'\\
&=\mathbf T\cdot\mathbf W'+\mathbf t\cdot\dot{\mathbf r}'-\mathbf t\cdot\mathbf W'\mathbf r-\mathbf t\cdot\mathbf W\mathbf r'\\
&=\mathbf M\cdot\mathbf W'+\mathbf t\cdot(\dot{\mathbf r}'-\mathbf W\mathbf r').
\end{aligned}
\end{equation}

\medskip

\subsection{Balance equations}
\color{black}As shown in the Appendix, t\color{black}he application of the \emph{principle of virtual powers} yields: 

\noindent 1) The \emph{configurational-balance equations}:
\begin{subequations}\label{eq:12}
  \begin{equation}
 S-\mathbf t\cdot\mathbf v-\mathbf T\cdot\mathbf V =B,\qquad \CC={\mathbf B_{\rm d}},
  \label{eq:13}
\end{equation}
where
\begin{equation}
  \label{eq:14}
\mathbf V=\mathbf R'\mathbf R^T\quad\textrm{and}\quad \mathbf v=\mathbf r'-\mathbf V\mathbf r.
\end{equation}
2) The \emph{standard-balance equations}:
\begin{align}
  \label{eq:15}
  &\mathbf T'+\mathbf B=\mathbf 0,\qquad \mathbf t'+\mathbf b=\mathbf 0.
\end{align}
\end{subequations}
3) The relation between referential and spatial stress descriptors:
\begin{equation}
  \label{eq:16}
\mathbf S=\mathbf R^T(\mathbf T-\mathbf r\wedge\mathbf t)\mathbf R.
\end{equation}
The first of \eqref{eq:13} can be reconciled with \eqref{eq:82} by defining 
\begin{align}
  \mathbf s=\mathbf R^T\mathbf t, 
\end{align}
and by observing that
\begin{align}\label{eq:84}
  \mathbf S\cdot\mathbf U+\mathbf s\cdot\mathbf u=\mathbf T\cdot\mathbf V+\mathbf t\cdot\mathbf v.
\end{align}

\subsection{Free energy and dissipation inequality}
The distorsion field determines, together with \emph{torsion-bending field} $\mathbf U$, the \emph{free energy per unit content}:
\begin{equation}
  \label{eq:10}
  \psi=\widehat\psi(\mathbf U-\UU),
\end{equation}
which we assume to have a stationary point at $\mathbf 0$. The field $\UU$ is to be interpreted as the torsion-bending field in a \emph{relaxed configuration} $\mathsf r_0=(\mathbf R_0,\mathbf r_0)$ obtained, \emph{modulo} an unessential rigid motion, by solving the differential equations $\mathbf R_0'=\mathbf R_0\UU$ and $\mathbf r_0'=\mathbf R_0^T\mathbf a$.\smallskip 

As a final step, we work out restrictions imposed on constitutive relations by the  dissipation inequality:
\begin{align}
  \frac{\rm d}{{\rm d}t}{\Psi}[(a,b)]\le\int_a^b \big(\mathbf S\cdot\dot{\mathbf U}+\CC\cdot\dotUU+S\mathring\lambda\big){\rm d}s,
\end{align}
where $\Psi[(a,b)]$, the total free energy contained in the part $(a,b)$, has been defined in \eqref{eq:4}. We deduce these restrictions by proceeding in the manner of Coleman-Noll \cite{ColemN1963ARMA}, bearing in mind the identity $$\frac{\rm d}{{\rm d}t}{\Psi}[(a,b)]=\int_a^b\big(\dot\psi+\psi \mathring\lambda\big){\rm d}s.$$
As a result, the system of stresses may be split as $\sigma=\sigma^{\rm ener}+\sigma^{\rm diss}$, where the energetic part $\sigma^{\rm ener}=(\mathbf S^{\rm ener},\CC^{\rm ener},S^{\rm ener})$ is defined by
\begin{subequations}
\begin{align}
  \label{eq:17}
  \mathbf S^{\rm ener}=\widehat\psi'(\mathbf U-\UU),\qquad \CC^{\rm ener}=-\widehat\psi'(\mathbf U-\UU),\qquad   S^{\rm ener}=\widehat\psi(\mathbf U-\UU).
\end{align}
and where the dissipative part $\sigma^{\rm diss}=(\mathbf S^{\rm diss},\CC^{\rm diss},N^{\rm diss})$ is required consistent with the residual dissipation inequality:
\begin{equation}
  \label{eq:18}
  \mathbf S^{\rm diss}\cdot\dot{\mathbf U}+\CC^{\rm diss}\cdot\dotUU+S^{\rm diss}\mathring\lambda\ge 0.
\end{equation}
\end{subequations}
In the next section we provide an example of constitutive prescriptions consistent with \eqref{eq:18}, and we construct two closed-form solutions displaying growth and remodeling.

\section{Specialization and examples}
In order to discuss concrete examples, we need to be more specific with our constitutive assumptions. To begin with, we assume that the \emph{free-energy mapping} $\widehat\psi(\cdot)$ is the sum of a constant $\psi_0$ and a positive-definite quadratic form:
\begin{subequations}\label{eq:21}
\begin{equation}
  \label{eq:22}
  \widehat\psi(\mathbf U_{\rm m})=\psi_0+\frac 12 \mathbb C\mathbf U_{\rm m}\cdot\mathbf U_{\rm m},\qquad\ \mathbf U_{\rm m}=\mathbf U-\UU,
\end{equation}
where $\psi_0$, the \emph{cost of axial accretion}, is  a positive constant. The linear map $\mathbb C$ transforms skew-symmetric tensors into skew-symmetric tensors. The simplest constitutive choice is:
\begin{equation}
  \label{eq:23}
  \mathbb C=GJ_t\mathbf A\otimes\mathbf A+EJ(\mathbb I-\mathbf A\otimes\mathbf A),
\end{equation}
\end{subequations}
where $G$ and $E$ are, respectively, the shear and Young's moduli, $J_t$ is the torsional moment of inertia, $J$ is the bending moment of inertia, $\mathbf A$ is the skew tensor associated to $\mathbf a$, and $\mathbb I$ is the identity map. Here the tensor product $\otimes$ between two skew-symmetric tensors $\mathbf A$ and $\mathbf B$ is defined by $(\mathbf A\otimes\mathbf B)\mathbf C=(\mathbf B\cdot\mathbf C)\mathbf A$. By \eqref{eq:21}, the first two equations in \eqref{eq:17} specialize into linear relations:
\begin{align}
  \label{eq:24}
  &\mathbf S^{\rm ener}=\mathbb C(\mathbf U-\UU),\qquad \CC^{\rm ener}=\mathbb C(\UU-\mathbf U).
\end{align}
For the dissipative part of the stresses, we select
\begin{equation}
  \label{eq:25}
  \mathbf S^{\rm diss}=\mathbf 0,\qquad\CC^{\rm diss}=\mathbb D\dotUU,\qquad S^{\rm diss}=\beta\mathring\lambda,
\end{equation}
where $\beta>0$ and
\begin{equation}
  \label{eq:26}
  \mathbb D=\tau_t GJ_t\mathbf A\otimes\mathbf A+\tau EJ(\mathbb I-\mathbf A\otimes\mathbf A),
\end{equation}
with $\tau_t$ and $\tau$ appropriate characteristic times. For the reader's purpose, we collect all relevant unknowns and equations in two boxes at the end of this section. 

\subsection{Example 1}
We suppose that the rod be clamped at $x=0$ and pulled by an axial force applied at $x=1$, as illustrated below.
\begin{figure}[H]
  \centering
\includegraphics[scale=0.5]{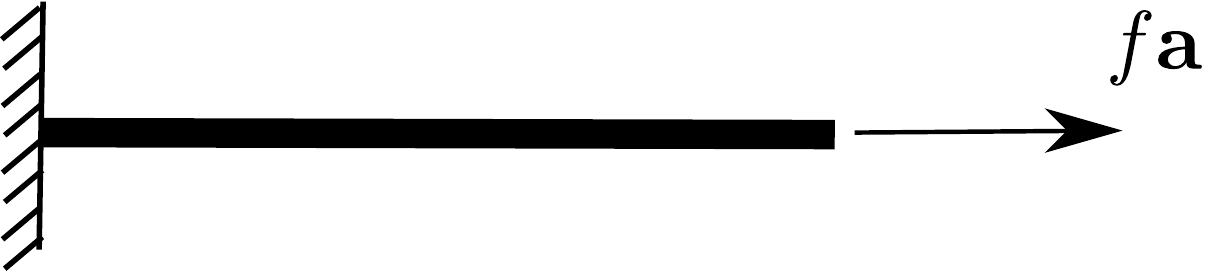}
  \caption{\footnotesize Rod pulled by an axial force}\label{fig:1}
\end{figure}
\noindent We assume that:
1) the relaxed strain $\UU$ vanishes at time $t=0$, that is, 
\begin{align}
  \label{eq:46}
&\UU(x,0)=\mathbf 0;
\end{align}
2) that standard and non-standard distributed external forces vanish, \emph{i.e.},
\begin{equation}
  \label{eq:48}
  \mathbf B=0,\qquad\mathbf b=0,\qquad {\mathbf B_{\rm d}}=0,\qquad B=0.
\end{equation}
We are going to show that the \emph{length} $\ell(t)$ of the rod obeys the exponential law
\begin{align}\label{eq:85}
  \ell(t)=\ell(0) \exp\Big(\frac{f-\psi_0}{\beta}t\Big),
\end{align}
where $\beta>0$ is the kinetic modulus introduced in \eqref{eq:35}. 

In order to derive \eqref{eq:85}, we begin by asking:
\begin{subequations}\label{eq:43}
  \begin{align}
  \label{eq:44}
  &\mathbf r(0,t)=\mathbf 0,\qquad\mathbf R(0,t)=\mathbf 0,\\
  &\mathbf t(1,t)=f\mathbf a,\qquad\mathbf T(1,t)=\mathbf 0,
\end{align}
\end{subequations}
in order to account for the imposed constraints. Then, we look for solutions of \eqref{eq:31}--\eqref{eq:42} such that
\begin{equation}
  \label{eq:49}
  \mathbf R=\mathbf I,\qquad\UU=\mathbf 0.
\end{equation}
Granted \eqref{eq:49}, we have ${\mathbf U}=\mathbf 0$. Hence, by \eqref{eq:37},
the rotational parts of the left stress and of the remodeling couple vanish:
\begin{equation}
  \label{eq:50}
  \mathbf T=\mathbf S=\mathbf 0,\qquad\CC=\mathbf 0.
\end{equation}
Moreover, we have
\begin{equation}\label{eq:51}
\mathbf t=f\mathbf a.
  \end{equation}
By substituting \eqref{eq:50} and \eqref{eq:51} in the constitutive equation \eqref{eq:41}, and by making use of \emph{Ansatz} \eqref{eq:49} we obtain $S=\psi_0-f+\beta\mathring\lambda$. Then, using \eqref{eq:48} and the balance equation \eqref{eq:34}, we arrive at the following evolution equation for $\lambda$:
\begin{equation}
\label{eq:52}
  \beta\frac{\dot{\lambda}}\lambda+\psi_0-f=0.
\end{equation}
The solution of \eqref{eq:52} is $\lambda(x,t)=\lambda(x,0)\exp\Big(\frac{f-\psi_0}{\beta}t\Big)$. Since $\ell(t)=\int_0^1 \lambda(x,t){\rm d}x$, we obtain \eqref{eq:85}.

\color{black}From this example, a flaw of our simple model becomes apparent: no growing body can reach a steady state when loaded by an external force, unless $f=\psi_0$, that is, the external force balances the cost of axial accretion. This feature appears  also in the theory of three-dimensional bulk growth, as pointed out in the final remarks of \cite{AmbroG2007MMS}. 

As a remedy, we may think of  replacing the last of \eqref{eq:25}, namely, the constitutive equation for the dissipative part of $S$, with a prescription drawn from the theory of viscoplasticity:
\begin{equation}
  \label{eq:96}
\begin{aligned}
  &S^{\rm diss}\in[-\gamma,+\gamma]&\textrm{if }\mathring\lambda=0,\\
  &S^{\rm diss}=\beta\mathring\lambda+\gamma\frac{\mathring\lambda}{|\mathring\lambda|}&\textrm{if }\mathring\lambda\neq 0,
\end{aligned}
\end{equation}
which is equivalent to asking that  $S^{\rm diss}$ be in the subdifferential of the dissipation potential $\zeta(\mathring\lambda)=\gamma|\mathring\lambda|+\frac 12 \beta\mathring\lambda^2$. Then, as is not difficult to check, the body would attain a steady state whenever $|f-\psi_0|\le\gamma$. Be as it may, the reader should bear in mind that the constitutive assumptions of this section have been selected only because of their extreme simplicity, and are not intended to be exhaustive of all possible behaviors. 
\color{black}

\subsection{Example 2}
We consider an initially-straight rod, whose terminal sections are instantaneously joined at time $t=0$, as shown in Fig. 5.
\begin{figure}[H]\label{fig:2}
  \centering
\def\svgwidth{0.6\linewidth}
%% Creator: Inkscape inkscape 0.48.4, www.inkscape.org
%% Accompanies image file 'ring.pdf' (pdf, eps, ps)
\begingroup%
  \makeatletter%
  \providecommand\color[2][]{%
    \errmessage{(Inkscape) Color is used for the text in Inkscape, but the package 'color.sty' is not loaded}%
    \renewcommand\color[2][]{}%
  }%
  \providecommand\transparent[1]{%
    \errmessage{(Inkscape) Transparency is used (non-zero) for the text in Inkscape, but the package 'transparent.sty' is not loaded}%
    \renewcommand\transparent[1]{}%
  }%
  \providecommand\rotatebox[2]{#2}%
  \ifx\svgwidth\undefined%
    \setlength{\unitlength}{408.39850464bp}%
    \ifx\svgscale\undefined%
      \relax%
    \else%
      \setlength{\unitlength}{\unitlength * \real{\svgscale}}%
    \fi%
  \else%
    \setlength{\unitlength}{\svgwidth}%
  \fi%
  \global\let\svgwidth\undefined%
  \global\let\svgscale\undefined%
  \makeatother%
  \begin{picture}(1,0.32077554)%
    \put(0,0){\includegraphics[width=\unitlength]{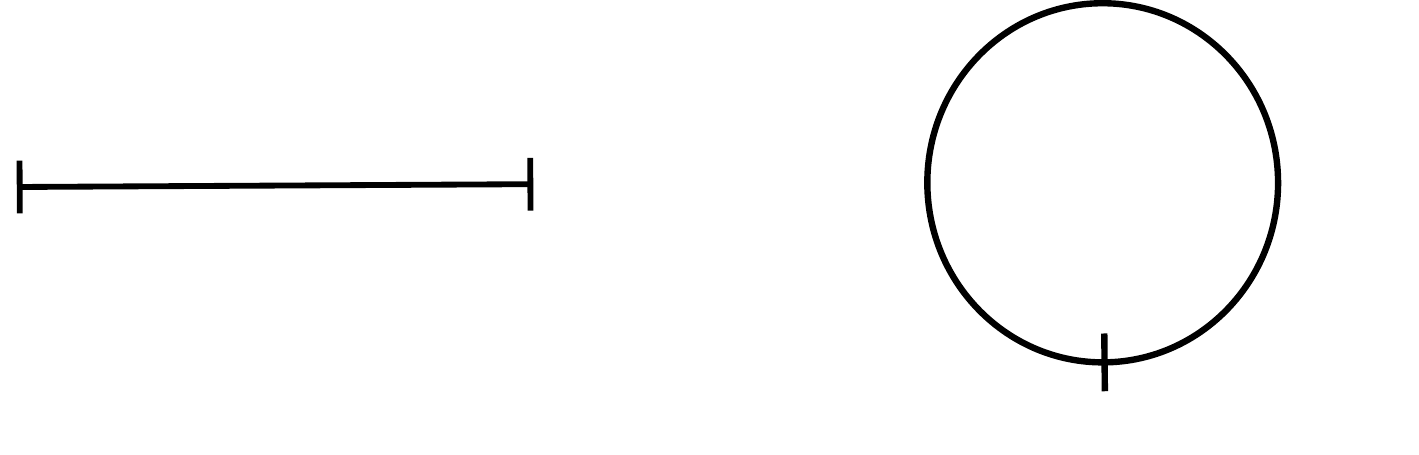}}%
    \put(-0.00195122,0.00488953){\color[rgb]{0,0,0}\makebox(0,0)[lb]{\smash{$\textrm{straight configuration}$}}}%
    \put(0.58771886,0.00686824){\color[rgb]{0,0,0}\makebox(0,0)[lb]{\smash{$\textrm{rolled-up configuration}$}}}%
  \end{picture}%
\endgroup%
\caption{\footnotesize A ring.}
\end{figure}
\noindent Accordingly, we impose
\begin{equation}
  \label{eq:55}
  \mathbf r(0,t)=\mathbf r(1,t),\qquad \mathbf R(0,t)=\mathbf R(1,t)
\end{equation}
for all $t\ge 0$. Moreover, as initial condition for $\mathbf U_{\rm d}$ we choose \eqref{eq:46}, and we suppose that the specific content at time $t=0$ be uniform:
\begin{align}
  \lambda(x,0)=\lambda_0.
\end{align}
We look for solutions of  \eqref{eq:31}--\eqref{eq:29} consistent with the following \emph{Ansatz}:
\begin{align}\label{eq:56}
  {\mathbf U}(x,t)=\kappa(t)\mathbf N,\qquad \UU(x,t)=\kappa_0(t)\mathbf N,\qquad \lambda(x,t)=\ell(t),\qquad\mathbf t=0,
\end{align}
where $\mathbf N$ is a skew tensor of unit norm such that $\mathbf A\cdot\mathbf N=0$. In order for \eqref{eq:55} to be satisfied, the length $\ell(t)$ and the curvature $\kappa(t)$ must comply with:
\begin{align}
\label{eq:57}
  \ell(t)=\frac {2\pi}{\kappa(t)}
\end{align}
at every time $t$. By combining \eqref{eq:56} with the constitutive equations \eqref{eq:37}, we obtain
\begin{subequations}\label{eq:58}
\begin{align}
  \label{eq:59}&\mathbf S=EJ(\kappa-\kappa_0)\mathbf N\\
  \label{eq:60}&\CC=EJ(\kappa_0-\kappa)\mathbf N+\tau EJ\dot\kappa_0\mathbf N,\\
  \label{eq:61}&S=\psi_0-\frac 1 2EJ(\kappa^2-\kappa_0^2)+\beta\dot\ell/\ell.
\end{align} 
\end{subequations}
A substitution of \eqref{eq:59} into the balance equations \eqref{eq:42} reveals that the standard-force balance \eqref{eq:32} is identically satisfied. On substituting (\ref{eq:58}b,c) into the remaining balance equations, taking into account \eqref{eq:57}, \eqref{eq:48} and \eqref{eq:49}, 
 we obtain the following system:
 \begin{align}\label{eq:65}
  & {\bar\tau}\dot\kappa_0=\frac{\kappa-\kappa_0}\alpha,\qquad \bar\tau\frac{\dot{\kappa}}\kappa=1-\frac {\kappa^2-\kappa_0^2}{\bar \kappa^2},
 \end{align}
where $\displaystyle{\bar\tau=\frac{\beta}{\psi_0}}$, and $\displaystyle{\bar\kappa=\sqrt{\frac{2\psi_0}{EJ}}}$ represent, respectively, a \emph{characteristic time} and a \emph{characteristic curvature}, and  $\displaystyle{\alpha=\frac{\bar\tau}\tau}$ is a parameter.

The characteristic time $\bar\tau$, and the characteristic curvature $\bar\kappa$ can be disposed of by a suitable change of scale in the dependent and independent variables. Indeed, on introducing the dimensionless variables $x_1=\displaystyle{\frac{\kappa_0}{\tilde\kappa}}$ and $x_2=\displaystyle{\frac{\kappa}{\tilde\kappa}}$, and on performing the change of timescale $\displaystyle{t\mapsto\frac t{\bar\tau}}$, we can write \eqref{eq:65} as
\begin{subequations}\label{eq:62}
\begin{align}\label{eq:69}
  &\dot x_1=\frac{x_2-x_1}\alpha,\qquad \dot x_2=(1+x_1^2-x_2^2)x_2.
\end{align}
By  supplementing \eqref{eq:69} with the initial conditions
\begin{align}\label{eq:70}
  x_1(0)=0,\qquad x_2(0)=\sqrt\frac{EJ}{2\psi_0}\frac{2\pi}{\lambda_0},
\end{align}
\end{subequations}
we obtain a standard Cauchy problem.

For $\alpha>0$, the function $\bm x\mapsto \begin{pmatrix}\frac{x_2-x_1}{\alpha}\\ (1+x_1^2-x_2^2)x_2 \end{pmatrix}$ is locally Lipschitz continuous, therefore, by the Picard--Lindel\"of theorem, the Cauchy problem \eqref{eq:62} has unique solution in some non-empty interval $[0,T)$. Moreover, $\dot x_2>0$ if $0<x_2<1$. Thus, given that $x_2(0)>0$, there exists $c>0$ such that $x_2(t)\ge c$ for all $t\in[0,T)$. As a consequence, $x_1(t)>0$ for all $t>0$. 

It is not difficult to prove that the solution can be prolonged up to an arbitrary time. To this aim, we notice that
\begin{equation}
  \label{eq:67}
   x_1^2(t)-x_2^2(t)\le 0\qquad\forall t\ge 0.
\end{equation}
Indeed, let us set $z=x_1^2-x_2^2$. The initial conditions \eqref{eq:70} entail that $z(0)<0$. Moreover, $z(t)$ can never become positive. In fact, by \eqref{eq:69}, we have $\dot z=2 x_1\frac{x_2-x_1}\alpha-2x_2^2(1+z)$, and hence $z=0$ implies $\dot z<0$. Once \eqref{eq:67} is established, it is immediately seen that $x_2(t)$ is a subsolution of $\dot y=y$, hence there exists a positive constant $C$ such that
\begin{equation}
  \label{eq:71}
  x_2(t)\le Ce^t\qquad\forall t\ge 0.
\end{equation}
As the growth rate of $x_1(t)$ does not exceed that of $x_2(t)$, we have
\begin{equation}
  \label{eq:79}
  x_1(t)\le Ce^t.
\end{equation}
Thanks to \eqref{eq:71} and \eqref{eq:79}, we conclude that the solution does not blow up in finite time. Consequently, standard continuation arguments for ODEs apply, and we conclude that the solution exists for all positive times.

We do not carry over the rigorous treatment of the qualitative behavior of the solutions of \eqref{eq:62}. Instead, we base our discussion on the inspection of the phase portrait displayed below, which we have drawn numerically choosing $\alpha=1$. 
\begin{figure}[H]
  \centering
\def\svgwidth{0.4\linewidth}
%% Creator: Inkscape inkscape 0.48.4, www.inkscape.org
%% Accompanies image file 'streamplot-of-the-solutions-of-the-ODE-for-the-ring.pdf' (pdf, eps, ps)
\begingroup%
  \makeatletter%
  \providecommand\color[2][]{%
    \errmessage{(Inkscape) Color is used for the text in Inkscape, but the package 'color.sty' is not loaded}%
    \renewcommand\color[2][]{}%
  }%
  \providecommand\transparent[1]{%
    \errmessage{(Inkscape) Transparency is used (non-zero) for the text in Inkscape, but the package 'transparent.sty' is not loaded}%
    \renewcommand\transparent[1]{}%
  }%
  \providecommand\rotatebox[2]{#2}%
  \ifx\svgwidth\undefined%
    \setlength{\unitlength}{663bp}%
    \ifx\svgscale\undefined%
      \relax%
    \else%
      \setlength{\unitlength}{\unitlength * \real{\svgscale}}%
    \fi%
  \else%
    \setlength{\unitlength}{\svgwidth}%
  \fi%
  \global\let\svgwidth\undefined%
  \global\let\svgscale\undefined%
  \makeatother%
  \begin{picture}(1,1)%
    \put(0,0){\includegraphics[width=\unitlength]{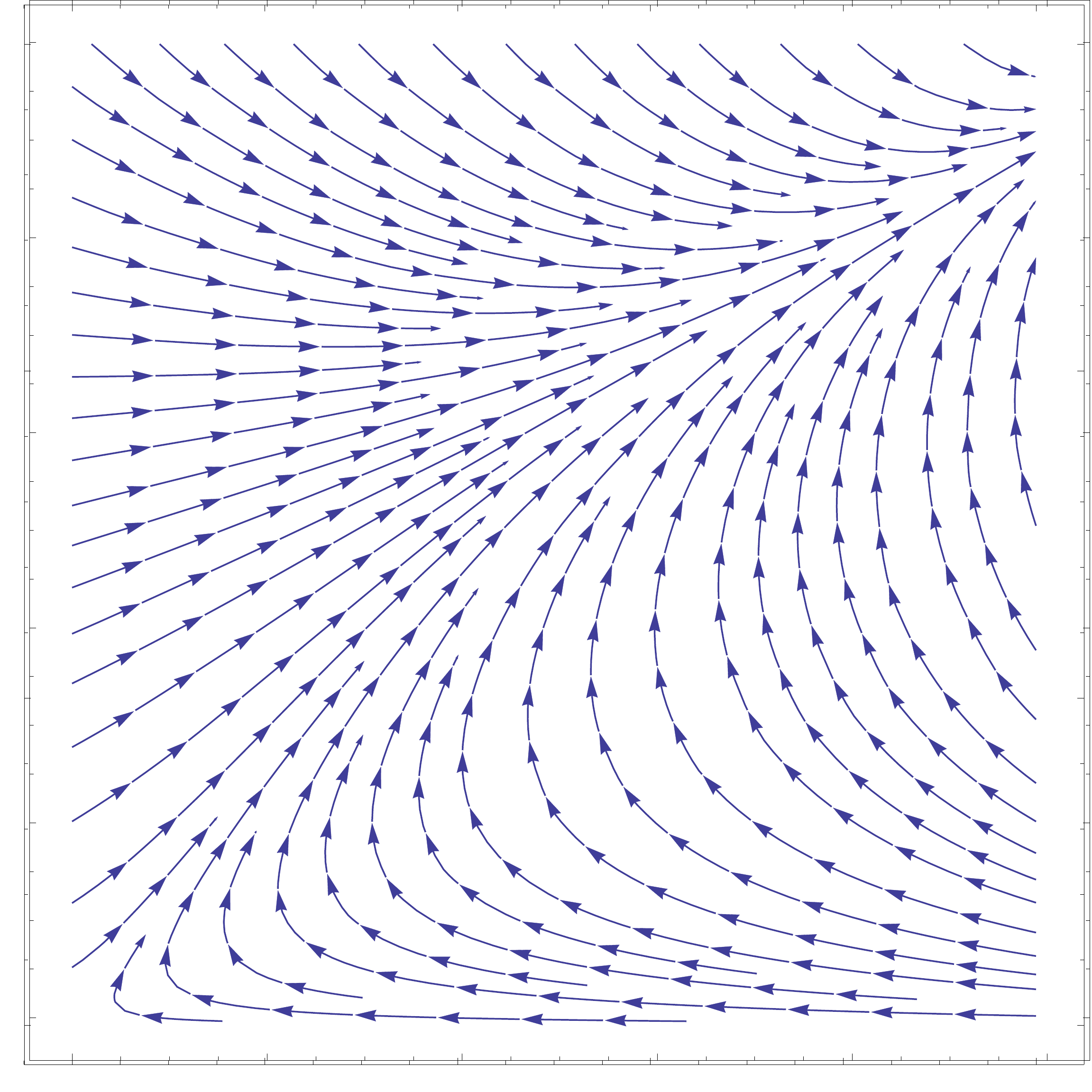}}%
    \put(0.47083334,-0.02588613){\color[rgb]{0,0,0}\makebox(0,0)[lb]{\smash{$\frac {\kappa_0}{\bar\kappa}$}}}%
    \put(-0.025,0.50625){\color[rgb]{0,0,0}\makebox(0,0)[lb]{\smash{$\frac{\kappa}{\bar\kappa}$}}}%
  \end{picture}%
\endgroup%
  \footnotesize\caption{Phase portrait of \eqref{eq:62} for $\displaystyle{\Big(\frac{\kappa_0}{\bar\kappa},\frac{\kappa}{\bar\kappa}\Big)\in(0,1)\times(0,3/2)}$.}
  \label{fig:11}
\end{figure}
The phase portrait in Figure 6 shows that if $\kappa(0)\ge\bar\kappa$, then  $x_2(t)=\frac{\kappa(t)}{\bar\kappa}$ is monotone increasing. If, instead, $\kappa(0)<\bar\kappa$, then $\kappa(t)$ decreases to a minimum, and then monotonically increases. In both cases, the curvature $\kappa(t)$ tends to $+\infty$ as $t\to+\infty$, hence the ring eventually shrinks to a point. This result is not surprising: taking into account the \emph{Ansatz} \eqref{eq:56}, and recalling \eqref{eq:57}, it is immedately seen that the amount of energy stored in the ring is
\begin{equation}
  \label{eq:72}
  \Psi=\frac{2\pi}{\kappa}\Big(\psi_0+\frac 12 EJ({\kappa-\kappa_0})^2\Big).
\end{equation}
During the first stage of the evolution process,  $\kappa_0$ is  small, and hence
\begin{equation}\label{eq:73}
\Psi\simeq \frac{\psi_0}\kappa+\frac 12 EJ\kappa.
\end{equation}
The  \emph{addenda} appearing on the right-hand side of \eqref{eq:73} are in competition: If $\kappa<\bar\kappa$, that is to say, $\frac 12 EJ\kappa^2<\psi_0$, then the \emph{addendum} $\frac {\psi_0}\kappa$ is dominant and energy minimization will drive $\kappa$ towards $+\infty$, so that the length (and hence the total content) tends to null. If $\kappa>\bar\kappa$, that is,  $\psi_0<\frac 12 EJ\kappa^2$, then the bending energy is dominant and, in order to reduce the total energy, the curvature of the ring must decrease. 

At subsequent stages of the evolution process, the spontaneous curvature $\kappa_0$ approaches $\kappa$, so that bending energy is negligible. In this case,
\begin{equation}
  \label{eq:78}
  \Psi\simeq \frac {\psi_0}\kappa,
\end{equation}
and energy minimization drives $\kappa$ towards $+\infty$, so that the length of the ring tends to null.

% \color{black}\section{Example 3}
% In this last example, in order to illustrate another example of \emph{stress relaxation}, we drop the hypothesis that the mechanical stretch  $\nu$ in the decomposition \eqref{eq:76} be equal to 1. The setting is the same as in Example 1, except that both ends of the rod are clamped. Let us assume that, at time $t=0$ the specific content be constant and equal to $\lambda_0$:
% \begin{align}
%   \lambda(0)=\lambda_0.
% \end{align}
% Immediately after time $t=0$, a mechanical stretch $\nu_0>1$ is impressed by suddenly increasing the spacing between the constraints. At each time $t>0$, the product of the growth stretch and mechanical stretch satisfies: 
% \begin{align}
%   \lambda(t)\nu(t)=\lambda_0\nu_0.
% \end{align}
% On denoting by $\mathbf a$ the unit vector parallel to the axis (see figure in Example 1) we let the axial component of the internal force $\mathbf t\cdot\mathbf a$ be related to the mechanical stretch $\nu$ by the constitutive equation:
% \begin{align}
%   \mathbf t\cdot\mathbf a=EA(\nu-1)=EA\Big(\frac{\nu_0\lambda_0}\lambda-1\Big)
% \end{align}
% where $A$ is the area of the cross section. Then, the evolution equation for the growth stretch is $ \beta \frac{\dot\lambda}\lambda+\psi_0-EA\Big(\frac{\nu_0\lambda_0}\lambda-1\Big)\frac {\lambda_0\nu_0} \lambda=B$, whence
% \begin{equation}
%   {\dot\lambda}+\lambda\frac{\psi_0-B}\beta=\frac{\nu_0EA}\beta
% \end{equation}
% \color{black}

\subsection{Summary of unknowns and governing equations}
\begin{center}
\framebox{
\begin{minipage}{0.8\textwidth}

\begin{center}
Box 1: List of unknowns
\vskip 1em

\footnotesize
Motion descriptors:
  \begin{equation*}
    \label{eq:27}
    \mathbf r(x,t)\in\mathcal V,\qquad\mathbf R(x,t)\in{\rm Rot}.
  \end{equation*}
State variables:
  \begin{align*}
    \label{eq:28}
    \mathbf U(x,t)\in{\rm Skw},\qquad\UU(x,t)\in{\rm Skw},
%  \end{align*}
 % \begin{equation*}
  \qquad  \lambda(x,t)\in\mathbb R^+.
  \end{align*}
Dynamical descriptors:
  \begin{equation*}
\mathbf T(x,t)\in{\rm Skw},\quad\mathbf t(x,t)\in\mathcal V,
%\end{equation*}
%\begin{equation*}
 \qquad   \mathbf S(x,t)\in{\rm Skw},\qquad\CC(x,t)\in{\rm Skw},
%\end{equation*}
%\begin{equation*}
\qquad S(x,t)\in\mathbb R.
\end{equation*}
\vskip 1em
$\mathcal V$: space of translation vectors\\
${\rm Skw}$: space of skew-symmetric tensors\\
${\rm Rot}$: special orthogonal group\\
${\mathbb R^+}$: positive real numbers\\[1em]
\null

\end{center}
\end{minipage}}

\framebox{
\begin{minipage}{0.45\textwidth}
\begin{center}
Box 2: governing equations.
\vskip 1em
\footnotesize
Balance:
\begin{subequations}\label{eq:31}
  \begin{align}
   \label{eq:32}&\mathbf T'+\mathbf B=\mathbf 0,\qquad \mathbf t'+\mathbf b=\mathbf 0\\
   \label{eq:33}&\CC={\mathbf B_{\rm d}}\\
   \label{eq:34}&S-\mathbf T\cdot\mathbf V-\mathbf t\cdot\mathbf v=B.
  \end{align}
\end{subequations}
Compatibility:
  \begin{align}\label{eq:35}
    &\mathbf U=\mathbf R^T\mathbf R',\qquad \mathbf u=\mathbf R^T\mathbf r'\\
    &\mathbf V=\mathbf R'\mathbf R^T,\qquad \mathbf v=\mathbf r'-\mathbf V\mathbf r.
\end{align}
Internal constraint:\vskip -1em
\begin{equation}
  \label{eq:36}
  \mathbf u(x,t)=\mathbf a.
\end{equation}
Response:
\begin{subequations}\label{eq:37}
  \begin{align}
    \label{eq:38}&\psi=\psi_0+\frac 12 \mathbb C(\mathbf U-\UU)\cdot(\mathbf U-\UU)\\
    \label{eq:39}&\mathbf S=\mathbb C(\mathbf U-\UU)\\
    \label{eq:40}&\CC=\mathbb C(\UU-\mathbf U)+\mathbb D\dotUU\\
    \label{eq:41}&S=\psi+\beta\mathring\lambda.
  \end{align}
\end{subequations}
Relation between $\mathbf T$ and $\mathbf S$:
  \begin{align}\label{eq:29}
    \mathbf S=\mathbf R^T(\mathbf T-\mathbf r\wedge\mathbf t)\mathbf R.
  \end{align}
Stiffness and viscosity tensors:
\begin{subequations}\label{eq:42}
  \begin{align}
 &\mathbb C=GJ_t\mathbf A\otimes\mathbf A+EJ(\mathbb I-\mathbf A\otimes\mathbf A),
\\
  & \mathbb D=\tau_t GJ_t\mathbf A\otimes\mathbf A+\tau EJ(\mathbb I-\mathbf A\otimes\mathbf A)\\
\nonumber&\textrm{where }\mathbf A\mathbf v=\mathbf a\times\mathbf v\quad\forall\mathbf v\in\mathcal V.
\end{align}
\end{subequations}
\end{center}
\end{minipage}
}
\end{center}

\color{black}
\section{Appendix}

\subsection{Three-dimensional bulk growth}
In this subsection we expound the main traits of the theory proposed in \cite{DiCarQ2002MRC}. Our treatment  slightly departs from \cite{DiCarQ2002MRC}, both in notation and in method, and is more close to \cite[Pt. XVI]{GurtiFA2010}, where the large-deformation theory of isotropic plasticity is discussed. This is not a coincidence, since this large-deformation plasticity has many points in common with the theory of bulk growth.  

The point of view guiding the development of the theory is that a continuum theory that aims at describing a certain class of bodies should be based on  a \emph{minimal set of balance principles} capturing the relevant physics, and valid for all instances of that class. The difference between material response within the same class should be incorporated into constitutive prescriptions. This point of view allows a systematic derivation of theories describing the  behavior of solids whose material structure may evolve, such as for instance plastic flow, phase transformation in shape-memory alloys and ferromagnetic materials.

Due to lack of intuition, it is not always clear what are the appropriate balance principles to be adopted. To this aim, the \emph{method of virtual powers} has proven to be extremely useful, as illustrated to a considerable extent in the recent monographs \cite{GurtiFA2010} and \cite{Fremo2012}. In order to be applied, this method requires, for each degree of freedom, a specification of power expenditure associated to its evolution. In particular, for a body undergoing bulk growth, it seems appropriate to assume that the external and internal powers have the form:
\begin{equation}\label{eq:126}
\begin{aligned}
  &\mathcal W_{\rm int}(\mathcal P)=\int_{\mathcal P} \bm{\mathbf T}_{\rm R}:\nabla \dot{\boldsymbol\chi}+J\bm{\mathbf C}:{\mathring{\mathbf  G}},\qquad {\rm and}
\\
& \mathcal W_{\rm ext}(\mathcal P)=\int_{\mathcal P} \bm{\mathbf b}_{\rm R}\cdot\dot{\boldsymbol\chi}+J\bm{\mathbf B}:{\mathring{\mathbf  G}}+\int_{\partial\mathcal P}\mathbf  t_{\rm R}(\mathbf  n)\cdot\dot{\boldsymbol\chi},
\end{aligned}
\end{equation}
Here, as usual in continuum mechanics, we describe the visible motion (\emph{i.e.}, the motion that may be observed at the macroscopic level) of a body by fixing once and for all a \emph{referential region} $\mathcal B$ and by considering a placement map $\boldsymbol\chi(\mathbf X,t)$ which associates to each $\mathbf X\in\mathcal B$ its position at time $t$. The domain of integration in \eqref{eq:126} is the typical \emph{referential body part} $\mathcal P\subset\mathcal B$. A double dot denotes standard contraction between tensors, that is, $\mathbf A:\mathbf B={\rm tr}(\mathbf A^T\mathbf B)$.

The dot products between the fields $\bm{\mathbf b}_{\rm R}$ and $\mathbf  t_{\rm R}(\mathbf  n)$ with the \emph{visible velocity} $\dot{\boldsymbol\chi}$ yield power densities per unit reference volume and per unit reference area, respectively. Since there is substantial agreement on notions such as velocity and mechanical power, one readily interprets $\mathbf  b_{\rm R}$ and $\mathbf  t_{\rm R}(\mathbf  n)$, respectively, as the \emph{referential body force density} and the \emph{referential traction density}. The latter accounts for contact inteactions between parts of the body, and between the body and its exterior; as such, is assumed  to depend only on the \emph{outward unit vector} $\mathbf  n$ to $\partial\mathcal P$, as usual in continuum mechanics. Likewise, one easily identifies the tensor field $\bm{\mathbf T}_{\rm R}$ with the \emph{Piola stress} \cite[\S 24.1]{GurtiFA2010}, since this field expends power on the \emph{referential gradient} $\nabla\dot{\boldsymbol\chi}$ of the visible velocity. 

The interpretation of the \emph{inner remodeling couple} $\bm{\mathbf C}$ and the \emph{outer remodeling couple} $\bm{\mathbf B}$ is an issue, since the notion of bulk growth as a \emph{microscopic process} is not part of common intuition. Yet, their introduction is mandatory if we are to account for power expenditure associated to the \emph{accretion rate}:
\begin{align}
  {\mathring{\mathbf  G}}=\dot{\mathbf  G}{{\mathbf  G}}^{-1},
\end{align}
a power whose density we renormalize with respect to the \emph{relaxed Jacobian}:
\begin{equation}
  \label{eq:105}
  J={\rm det}\mathbf  G
\end{equation}
which delivers volume change associated to accretion. A link between accretive forces and homeostatic stress has been established in \cite{AmbroG2007MMS}; however, this interpretation can be given only \emph{a posteriori}, once a model has been assembled and its predictions have been compared with physical observation. At this stage, we content ourselves with regarding accretion stresses as \emph{primitive objects}, just as standard forces.

Let us assume that \emph{external and internal powers be balanced}:
\begin{equation}
  \label{eq:100}
  \mathcal W_{\rm int}(\mathcal P)-\mathcal W_{\rm ext}(\mathcal P)=0
\end{equation}
for every pair $(\dot{\boldsymbol\chi},{\mathring{\mathbf  G}})$, regarded as \emph{arbitrary test functions}, and for each part $\mathcal P$. By the application of the divergence theorem, \eqref{eq:100} becomes:
\begin{equation}
  \label{eq:92}
  \int_{\mathcal P} ({\rm Div}\bm{\mathbf T}_{\rm R}+\mathbf  b_{\rm R})\cdot\dot{\boldsymbol\chi}+J(\bm{\mathbf  B}-\mathbf  C):\mathring{\mathbf  G}+\int_{\partial\mathcal P}({\mathbf  t}_{\rm R}(\mathbf  n)-\mathbf  T_{\rm R}\mathbf  n)\cdot\dot{\boldsymbol\chi}=0.
\end{equation}
By choosing virtual velocities that vanish on $\partial\mathcal P$, and such that $\dot{\mathbf  G}=0$, and by a localization argument (see \cite{GurtiFA2010}), we obtain the \emph{standard-force balance}:
\begin{align}\label{eq:54}
{\rm div}\bm{\mathbf T}_{\rm R}+\bm{\mathbf b}_{\rm R}=\bm{\mathbf 0}.
\end{align}
With \eqref{eq:54} at hand, taking velocities that do not necessarily vanish on $\partial\mathcal P$, and using again a localization argument we obtain the \emph{Cauchy relation}:
\begin{equation}
  \label{eq:99}
  \mathbf T_{\rm R}\mathbf  n=\mathbf  t_{\rm R}(\mathbf  n)
\end{equation}
between Piola stress and referential traction. The other consequence of the  principle of virtual powers, namely the \emph{balance of accretion forces}: 
\footnote{The reader should be willing to accept \eqref{eq:110} as a basic postulate, in the same way as she/he accepts  balance of forces without the need of having a force defined in terms of other objects. We illustrate this point of view with a very simple example: a spring with a hanging mass. The starting point to construct a model  delivering the  \emph{elongation} $x$ of the spring is the \emph{force balance}:
  \begin{equation}
    \label{eq:97}
     f_{\rm int}=f_{\rm ext}
  \end{equation}
involving  the \emph{external force} $f_{\rm ext}$ and the \emph{internal restoration force} $f_{\rm int}$ exerted by the spring. At this stage, we are anticipating that $x$ will be determined through an identity involving two new objects that are not defined. Yet, nobody would reject \eqref{eq:9}, being used to forces as \emph{primitive} construct, and to force balance as a postulate of mechanics. The elongation $x$ would then appear in the model once we augment the balance equation with appropriate constitutive prescriptions. For instance:
\begin{equation}
  \label{eq:125}
  f_{\rm int}=kx,\qquad  \textrm{and}\qquad f_{\rm ext}=mg
\end{equation}
with $k$ the \emph{stiffness} of the spring, $m$ the \emph{mass}, and  $g$ the \emph{gravitational acceleration constant}. 
}
\begin{align}\label{eq:110}
  \mathbf  C=\bm{\mathbf  B},
\end{align}
is arrived at by localization, on letting $\dot{\mathbf  G}$ be arbitrary. 

When looking at \eqref{eq:54} and \eqref{eq:110}, one cannot but notice that there is no coupling between the balance laws governing visible motion and material remodeling. Such coupling is indeed \emph{constitutive}: it appears when \eqref{eq:54} and \eqref{eq:110} are augmented with suitable constitutive specifications relating stress and remodeling couples to the actual motion of the body. Although there is a great deal of freedom in the choice these specifications, they must be  consistent with the following dissipation principle: \emph{the rate of change of free-energy $\Phi(\mathcal P)$ stored in a part $\mathcal P$ must not exceed the power supplied on that part by its exterior.} Since external and internal power expenditures are equal, this statement is equivalent to:
\begin{equation}
  \label{eq:101}
  \frac{\rm d}{{\rm d}t}\Phi(\mathcal P)\le \mathcal W_{\rm int}(\mathcal P).
\end{equation}
What distinguishes bulk growth from standard elasticity is that the free energy density per unit referential volume, besides depending on the mechanical distortion $\mathbf  F$,  is proportional to the relaxed Jacobian $J$ defined in \eqref{eq:105}. Accordingly, the free energy stored in $\mathcal P$ is:
\begin{equation}
  \label{eq:102}
   \Phi(\mathcal P)=\int_{\mathcal P}J\check\varphi(\mathbf  F)
\end{equation}
From \eqref{eq:102}, using the identity $\dot J=J\mathbf  G^{-T}\cdot\dot{\mathbf  G}$ and a localization argument, we obtain the inequality:
\begin{equation}\label{eq:106}
  J{\mathbf  G}^{-T}\cdot\dot{\mathbf  G}\check\varphi(\mathbf  F)+J\partial\check\varphi(\mathbf  F)\cdot\dot{\mathbf  F}\le \mathbf  T_{\rm R}\mathbf  G^{T}\cdot\dot{\mathbf  F}+{\mathbf  F}^{T}\mathbf  T_{\rm R}\cdot\dot{\mathbf  G}+J\mathbf  C\cdot\dot{\mathbf  G}\mathbf  G^{-1}.
\end{equation}
Then, on introducing the new stresses:
\begin{align}
  \mathbf  S=J^{-1}\mathbf  T_{\rm R} \mathbf  G^T,\qquad   \mathbf  M=\mathbf  F^T\mathbf  S,
\end{align}
and dividing both sides of \eqref{eq:106} by $J$, we obtain:
\begin{equation}\label{eq:103}
(\mathbf  S-\partial\check\varphi(\mathbf  F))\cdot\dot{\mathbf  F}+(\mathbf  C+\mathbf  M-\check\varphi(\mathbf  F)\mathbf  I)\cdot\mathring{\mathbf  G}\ge 0.
\end{equation}
On setting:
\begin{align}\label{eq:129}
{\mathbf  S}_+=\mathbf  S-\partial\varphi(\mathbf  F),\qquad {\mathbf  C}_+=\mathbf  C+\mathbf  M-\check\varphi(\mathbf  F)\mathbf  I,
\end{align}
the dissipation inequality takes the form:
\begin{align}\label{eq:108}
{\mathbf  S}_+\cdot\dot{\mathbf  F}+{\mathbf  C}_+\cdot\mathring{\mathbf  G}\ge 0.
\end{align}
Assume a constitutive dependence of ${\mathbf  S}_+$ and ${\mathbf  C}_+$ on the list $\Lambda=(\mathbf  F,\mathbf  G,\dot{\mathbf  F},\mathring{\mathbf  G})$ through functions  $\check{\mathbf  S}_+(\Lambda)$ and $\check{\mathbf  C}_+(\Lambda)$, namely,
\begin{equation}
  \label{eq:109}
  \mathbf  S_+=\check{\mathbf  S}_+(\mathbf  F,\mathbf  G,\dot{\mathbf  F},\mathring{\mathbf  G}),\qquad   \mathbf  C_+=\check{\mathbf  C}_+(\mathbf  F,\mathbf  G,\dot{\mathbf  F},\mathring{\mathbf  G}).
\end{equation}
Granted that these functions are \emph{smooth}, an argument in the Appendix of \cite{BertsPV2001AMPA} yields:
\begin{equation}
  \label{eq:107}
   \check{\mathbf  S}_+(\mathbf  F,\mathbf  G,\mathbf  0,\mathbf  0)=\mathbf  0,\qquad    \check{\mathbf  C}_+(\mathbf  F,\mathbf  G,\mathbf  0,\mathbf  0)=\mathbf  0.
\end{equation}
Indeed, let us fix ${\mathbf F}$ and ${\mathbf G}$, and let us assume that for every triplet $(\varepsilon,\mathbf V,\mathbf W)$ of a scalar $\varepsilon>0$ and tensors $\mathbf  V$ and $\mathbf  W$, it is possible to realize a process $t\mapsto(\mathbf F(t),\mathbf G(t))$ such that, at a certain time $\overline t$, $\mathbf F(\overline t)={\mathbf F}$, $\mathbf G(\overline t)={\mathbf G}$ and
\begin{equation}
  \label{eq:117}
  \dot{\mathbf  F}(\overline t)=\varepsilon\mathbf  V,\qquad \mathring{\mathbf  G}(\overline t)=\varepsilon\mathbf  W.
\end{equation}
Substituting into \eqref{eq:109}--\eqref{eq:107}, and dividing by $\varepsilon$, we obtain:
\begin{equation}
  \label{eq:118}
\check{\mathbf  S}_+({\mathbf  F},{\mathbf  G},\varepsilon\mathbf  V,\varepsilon\mathbf  W)\cdot\mathbf  V+\check{\mathbf  C}_+({\mathbf  F},{\mathbf  G},\varepsilon\mathbf  V,\varepsilon\mathbf  W)\cdot{\mathbf  W}\ge 0.
\end{equation}
On letting $\varepsilon\to 0$, we get:
\begin{equation}
  \label{eq:119}
\check{\mathbf  S}_+({\mathbf  F},{\mathbf  G},\mathbf  0,\mathbf  0)\cdot\mathbf  V+\check{\mathbf  C}_+({\mathbf  F},{\mathbf  G},\mathbf  0,\mathbf  0)\cdot{\mathbf  W}\ge 0.
\end{equation}
Since \eqref{eq:119} must hold for whatever choice of $\mathbf  V$ and $\mathbf  W$, we arrive at \eqref{eq:107}. 

Now, a consequence of  \eqref{eq:107} is that:
\begin{equation}
  \label{eq:114}
  \begin{aligned}
    \widehat{\mathbf  S}_+(\bullet)=\widehat{\mathbb K}_{(\mathbf F\mathbf F)}(\bullet)\dot{\mathbf  F}+\widehat{\mathbb K}_{(\mathbf  F\mathbf G)}(\bullet)\mathring{\mathbf  G},\\
\widehat{\mathbf  C}_+(\bullet)=\widehat{\mathbb K}_{(\mathbf G\mathbf F)}(\bullet)\dot{\mathbf  F}+\widehat{\mathbb K}_{(\mathbf G\mathbf G)}(\bullet)\mathring{\mathbf  G},
  \end{aligned}
\end{equation}
where $\mathbb K_{(\bullet\bullet)}(\mathbf  F,\mathbf  G,\dot{\mathbf  F},\mathring{\mathbf  G})$ stands for a linear map which transforms linearly tensors into tensors. Consistency with the dissipation principle requires that the block matrix:
\begin{equation}
  \label{eq:113}
 \widehat{\mathbb K}=\blockmatrix{\widehat{\mathbb K}_{\mathbf F\mathbf F}}{\widehat{\mathbb K}_{\mathbf F\mathbf G}}{\widehat{\mathbb K}_{\mathbf G\mathbf F}}{\widehat{\mathbb K}_{\mathbf G\mathbf G}}
\end{equation}
be non-negative; the simplest, yet nontrivial choice consistent with such requirement is:
\begin{equation}
  \label{eq:115}
  \widehat{\mathbb K}=\beta\blockmatrix{0}{0}{0}{\mathbb I},
\end{equation}
with $\beta>0$. Then, Piola stress and inner accretive remodeling couple are given, respectively, by
\begin{equation}
  \label{eq:116}
\mathbf  T_{\rm R}=J\partial\check\varphi(\mathbf  F)\mathbf  G^{-T},\qquad  \textrm{and}\qquad \mathbf  C=\mathbf  E+\beta\mathring{\mathbf  G},
\end{equation}
where\begin{align}\label{eq:74b}
  \bm{\mathbf E}=\widehat\varphi(\bm{\mathbf F})\,\bm{\mathbf I}-\bm{\mathbf F}^{\rm T}\color{black}\partial\widehat{\varphi}(\bm{\mathbf F}).\color{black} 
\end{align}
By combining the balance of accretion forces \eqref{eq:110} with the constitutive prescription $\eqref{eq:116}_2$ we obtain  the following evolution equation:
\begin{align}
  \beta\dot{\mathbf  G}\bm{\mathbf G}^{-1}+\widehat\varphi(\bm{\mathbf F})\,\bm{\mathbf I}-\bm{\mathbf F}^{\rm T}\partial\widehat{\varphi}(\bm{\mathbf F})=\bm{\mathbf B}.
\end{align}
As also discussed in \cite{AmbroG2007MMS}, the main point with \eqref{eq:74b} is that it identifies a key coupling mechanism between stress and growth irrespective of the choice of free-energy and dissipation. Of course, \emph{additional coupling mechanisms} may be introduced in the model: for instance, through the outer remodeling couple  $\bm{\mathbf B}$. Indeed the modifier \emph{``outer''}, for the remodeling couple $\bm{\mathbf B}$ refers to the fact that its working is not accounted for in the dissipation principle and hence stands outside the thermodynamic structure of the theory. Yet,  $\bm{\mathbf B}$ needs not be ascribed solely to agents outside the body, and may well depend on processes that take place \emph{within the body}.

The theory of bulk growth and the theory of plastic solids undergoing large deformations have many common traits. Both are based on the decomposition \eqref{eq:66}. However, in the latter theory $\mathbf  G$ is interpreted as the \emph{plastic-distortion tensor}.\footnote{In \cite{GurtiFA2010} the plastic distortion tensor is denoted by  $\mathbf F^{\rm p}$.} Since plastic flow is not accompanied by appreciable volume changes, plastic distortion is \emph{isochoric} \cite[Eq. (91.1)]{GurtiFA2010}:
\begin{equation}
  \label{eq:119b}
  J={\rm det}\mathbf  G=1.
\end{equation}
As a consequence $\mathring{\mathbf  G}$ is \emph{deviatoric}:
\begin{equation}
  \label{eq:121}
  {\rm tr}\mathring{\mathbf  G}=0,
\end{equation}
and a balance similar to that of accretion forces holds, but only for the deviatoric parts (here denoted with a subscript $0$ as in \cite{GurtiFA2010}) of the relevant microscopic stresses:
\begin{equation}
  \label{eq:124}
  \mathbf  C_0=\mathbf  B_0.
\end{equation}
More importantly, because of \eqref{eq:121}, the term $\check\varphi(\mathbf  F)$ disappears from the dissipation inequality, which then reads:
\begin{equation}\label{eq:88}
(\mathbf  S-\partial\check\varphi(\mathbf  F))\cdot\dot{\mathbf  F}+(\mathbf  C+\mathbf  M)\cdot\mathring{\mathbf  G}\ge 0.
\end{equation}
In particular, \eqref{eq:129} is replaced by
\begin{equation}
  \mathbf  C_{\rm +}=\mathbf  C-\mathbf  M.
\end{equation}
Thus, the balance equation \eqref{eq:124} can be written as:
\begin{equation}
  \mathbf  M_0=(\mathbf  C_+)_0+\mathbf  B_{\rm 0}.
\end{equation}
Notice that, on denoting by $\mathbf  T=(\nabla{\boldsymbol\chi})^{-1}\mathbf  T_{\rm R}(\nabla\boldsymbol\chi)^{T}$ the Cauchy stress, we have $\mathbf  M=\mathbf  F^T\mathbf  S=J^{-1}\mathbf  F^T\mathbf  T_{\rm R}\mathbf  G^T={\rm det}(\nabla{\boldsymbol\chi})\mathbf  F^T\mathbf  T(\nabla\boldsymbol\chi)^{-T}\mathbf  G^T={\rm det}(\nabla{\boldsymbol\chi})\mathbf  F^T\mathbf  T(\nabla\boldsymbol\chi)^{-T}\mathbf  G^T$, whence:\footnote{Since  ${\rm tr}(\mathbf  F^T\mathbf  T\mathbf  F^{-T})={\rm tr}(\mathbf  F^{-T}\mathbf  F^T\mathbf  T)={\rm tr}(\mathbf  T)$,
we have also $\mathbf  M_0=\mathbf  M-\frac 13 {\rm tr}(\mathbf  M)\mathbf  I=\mathbf  M- {\rm det}(\nabla{\boldsymbol\chi})\frac 13{\rm tr}(\mathbf  T)\mathbf  I={\rm det}(\nabla{\boldsymbol\chi})\mathbf  F^{T}(\mathbf  T-\frac 13{\rm tr}(\mathbf  T)\mathbf  I)\mathbf  F^{-T}$, whence
\begin{equation}
\begin{aligned}
  \mathbf  M_0={\rm det}(\nabla{\boldsymbol\chi})\mathbf  F^T\mathbf  T_0\mathbf  F^{-T}.
\end{aligned}
\end{equation}}
\begin{equation}
  \label{eq:127}
  \begin{aligned}
    \mathbf  M={\rm det}(\nabla{\boldsymbol\chi})\mathbf  F^T\mathbf  T\mathbf  F^{-T}.
  \end{aligned}
\end{equation}
Thus, by a comparison with Eq. (94.12) of \cite{GurtiFA2010}, we see that $\mathbf  M$ becomes the \emph{Mandel stress}, a tensorial quantity that, in finite plasticity, conveys the interaction between stress and plastic flow. 

The analogy between bulk growth and plastic distortion suggests two ways of modifying the thery of three-dimensional bulk growth presented in this section. First, one may replace the smooth constitutive equation for ${\mathbf  C}_+$ with an inclusion of the form:
\begin{equation}
  \label{eq:128}
{\mathbf  C}_+\in\partial\zeta(\mathring{\mathbf  G}),
\end{equation}
where $\zeta$ is a possibly \emph{nonsmooth} dissipation potential such that $0\in\partial\zeta(\mathbf  0)$, and $\zeta(\mathring{\mathbf  G})$ is its subdifferential set at $\mathring{\mathbf  G}$. Non-smooth dissipation potentials are customary in the mathematical modeling of materials that display hysteresis \cite{Fremo2002,Visin1994}, for instance, in the modeling of shape memory alloys \cite{RoubT2013ARMA,RoubT2010ZAMM} and ferromagnetic materials \cite{PodioRT2010ARMA,RoubT2011MMMAS}. Of course, the argument following \eqref{eq:109} would not apply in this case. Nevertheless, the dissipation inequality would still hold. Such modification would likely introduce hysteresis effects and, moreover, would provide a remedy to a limitation pointed out in \cite{AmbroG2007MMS}. Namely, that a growing body under constant external load never reaches a stationary configuration. 

Second, it is known that the inclusion of the gradient of plastic strain in the internal power  expenditure \cite{Gurti2004JMPS} leads to models that can replicate size dependent resistance to plastic flow observed for instance in torsion experiments \cite{FleckMAH1994AMM} as confirmed by recent analytical results \cite{ChiriGT2012SJAM,IdiarF2010MSMS}. 
We therefore argue that the inclusion, in the expression of the internal power, of a term accounting for power expenditure associated the gradient of the accretion velocity would lead to interesting size-dependent interaction between stress and growth.

\subsection{Right and left derivatives}
The application of the principle of virtual powers to derive balance equations appropriate to rod theories is substantially simplified by using machinery and concepts from the theory of Lie groups, as pointed out in \cite{NardiTT2002EJMA}.

As explained in Section 2.1, a rod is regared as a one-dimensional collection of sections, labeled through a parameter $x$ in an open interval which, for lack of a better choice, we take to be $(0,1)$. Accordingly, the motion of a rod may be represented through a function
\begin{align}\label{eq:45}
  (x,t)\mapsto \mathsf r(x,t)\in\mathcal G,
\end{align}
where $\mathcal G$ is the Lie Group (see for instance \cite[Chap. 12]{CrampP1986}) of \emph{rigid-body motions}, i.e., orientation-preserving isometries of the three-dimensional Euclidean point space. On choosing a pole $p_0$ and on orthonormal basis, one may think of any such isometry as a rotation keeping $p_0$ fixed, represented through a rotation matrix $\mathbf R(x,t)$, followed by a translation $\mathbf r(x,t)$, in the manner prescribed by \eqref{eq:1}.

When doing calculations on the group $\mathcal G$, it is useful to identify its typical element $\mathsf r$ with a $\mathbb R^{4\times 4}$ matrix, which in block-matrix notation can be rendered as:
\begin{align}\label{eq:90}
  \mathsf r=\blockmatrix{\mathbf R}{\mathbf r}{\mathbf 0}1,
\end{align}
so that composition between rigid-body motions can be computed by performing matrix multiplication. By doing so, one easily finds that the inverse of $\mathsf r$ has the representation:
\begin{align}
  \mathsf r^{-1}=\blockmatrix{\mathbf R^T}{-\mathbf R^T\mathbf r}{\mathbf 0}1.
\end{align}

The velocity of a particular section $x$ at a given time $t$:
\begin{align}
  \dot{\mathsf r}(x,t)=\blockmatrix{\dot{\mathbf R}(x,t)}{\dot{\mathbf r}(x,t)}{\mathbf 0}0
\end{align}
is an element of $T_{\mathsf r(x,t)}\mathcal G$, the tangent space at $\mathsf r(x,t)$. Consider now two sections, say $x_1$ and $x_2$. In general, we shall have $\mathsf r(x_1,t)\neq\mathsf r(x_2,t)$, thus the velocities of section $x_1$ and section $x_2$ are elements of distinct tangent spaces. In order to compare these velocities, we need to transport the corresponding tangent vectors in the tangent space of the same point. It is then natural to choose this point as the identity $\mathsf e=\blockmatrix{\mathbf I}{\mathbf 0}{\mathbf 0}1$. 

One method to pull the velocity vector $\dot{\mathsf r}$ back to the identity is by \emph{right-composition} with $\mathsf r^{-1}$. In order to define this operation, we consider the curve
\begin{align}
\nonumber\tau\mapsto&\mathsf r(x,t+\tau)\mathsf r^{-1}(x,t)\\
&=\blockmatrix{\mathbf R(x,t+\tau)\mathbf R^T(x,t)}{\mathbf r(x,t+\tau)-\mathbf R(x,t+\tau)\mathbf R^T(x,t)\mathbf r(x,t)}{\mathbf 0}1.
\end{align}
At $\tau=0$ this curve goes through $\mathsf e$. Accordingly, its derivative at $\tau=0$, which we refer to as the \emph{right velocity}:
\begin{align}
\nonumber
{\bm{\mathsf w}}(x,t)&=\left.\frac{\rm d}{{\rm d}\tau}\right|_{\tau=0}\!\!\!\!\!{\mathsf r}(x,t+\tau)\mathsf r^{-1}(x,t)
\\
\nonumber
&=\dot{\mathsf r}(x,t)\mathsf r^{-1}(x,t)
\\
\nonumber
&=\blockmatrix{\dot{\mathbf R}(x,t)\mathbf R^T(x,t)}{\dot{\mathbf r}(x,t)-\dot{\mathbf R}(x,t)\mathbf R^T(x,t)\mathbf r(x,t)}{\mathbf 0}0
\nonumber
\\
&=\blockmatrix{\mathbf W(x,t)}{\mathbf w(x,t)}{\mathbf 0}0
\end{align}
is an element of $T_{\mathsf e}\mathcal G$, the tangent space at $\mathsf e$. It is worth noticing that $\mathbf W$ is a skew-symmetric tensor (as can be verified by differentiating the identity $\mathbf R^T\mathbf R=\mathbf I$). Consequently, there exists a unique vector ${\bm\omega}$, the \emph{axial vector of $\mathbf W$}, such that ${\bm\omega}\times\mathbf a=\mathbf W\mathbf a$ for every vector $\mathbf a$. The  vectors ${\bm{\omega}}(x,t)$ and $\mathbf w(x,t)$ are, respectively, the \emph{spin} and the \emph{velocity} characterizing the rigid-velocity field of section $x$ at time $t$, resolved with respect to the fixed pole $p_0$. 

Drawing our terminology from the theory of Lie Groups, we say that the right velocity is the image of $\dot{\mathsf r}$ under linear map  $(R_{\mathsf r^{-1}})_*:T_{\mathsf r}\mathcal G\to T_{\mathsf e}\mathcal G$ induced by the \emph{right translation} $R_{\mathsf r^{-1}}:\mathcal G\to\mathcal G$, the map defined by 
$R_{\mathsf r^{-1}}\mathsf g=\mathsf g\mathsf r^{-1}$  (see \cite[Ch. 12 \S 5]{CrampP1986}). 

The notion of right derivative can be used also when differentiating the function \eqref{eq:45} with respect to $x$. Let us consider the curve:
\begin{align}\label{eq:86}
  \xi\mapsto & \mathsf r(x+\xi,t)\mathsf r^{-1}(x,t)\nonumber \\
\qquad &=\blockmatrix{\mathbf R(x+\xi,t)\mathbf R^T(x,t)}{\mathbf r(x+\xi,t)-\mathbf R(x+\xi,t)\mathbf R^T(x,t)\mathbf r(x,t)}{\mathbf 0}1.
\end{align}
Again, this curve goes through $\mathsf e$ at $\xi=0$. Hence, its derivative at $\xi=0$, namely,
\begin{align}
\nonumber
&\left.\frac{\rm d}{{\rm d}\xi}\right|_{\xi=0}\!\!\!\!\!
\mathsf r(x+\xi,t)\mathsf r^{-1}(x,t)
\\
\nonumber
&\qquad=(\partial_x\mathsf r(x,t))\mathsf r^{-1}(x,t)
\\
&\qquad=\blockmatrix{\partial_x{\mathbf R}(x,t)\mathbf R^T(x,t)}{\partial_x{\mathbf r}(x,t)-\partial_x{\mathbf R}(x,t)\mathbf R^T(x,t)\mathbf r(x,t)}{\mathbf 0}0
\end{align}
is a tangent vector at $\mathsf e$. If we replace differentiation with respect to $x$ with differentiation with respect to the content measure $s$, defined in \eqref{eq:26}, then we obtain the \emph{right strain}:
\begin{align}
\nonumber
  \mathds v&=\lambda^{-1}(\partial_x\mathsf r)\mathsf r^{-1}
\\
\nonumber
&=\mathsf r'\mathsf r^{-1}\\
\nonumber
&=\blockmatrix{\mathbf R'\mathbf R^T}{\mathbf r-\mathbf R'\mathbf R\mathbf r}{\mathbf 0}{0}
\\
&=\blockmatrix{\mathbf V}{\mathbf v}{\mathbf 0}{0}.
\end{align}

Another way to pull tangent vectors back to the identity is by \emph{left translation}. The curve:
\begin{align}\label{eq:87}
& \xi\mapsto \mathsf r^{-1}(x,t)\mathsf r(x+\xi,t)\nonumber
\\
&\qquad=\blockmatrix{\mathbf R^T(x,t)\mathbf R(x+\xi,t)}{-\mathbf R^T(x,t)\mathbf r(x,t)+\mathbf R(x,t)^T\mathbf r(x+\xi,t)}{\mathbf 0}1
\end{align}
goes through $\mathsf e$ at $\xi=0$. Its derivative at $\xi=0$:
\begin{align}
  &\left.\frac{\rm d}{{\rm d}\xi}\right|_{\xi=0}\!\!\!\!\!
\mathsf r^{-1}(x,t)\mathsf r(x+\xi,t)
\nonumber\\
\nonumber
&\qquad=\mathsf r^{-1}(x,t)\partial_x\mathsf r(x,t)
\\
\nonumber
&\qquad=\blockmatrix{{\mathbf R}^T(x,t)\partial_x\mathbf R(x,t)}{{\mathbf R}^T(x,t)\partial_x\mathbf r(x,t))}{\mathbf 0}0.
\end{align}
is a tangent vector at $\mathsf e$. We define the \emph{left strain}, 
\begin{align}
\nonumber
\mathds u&=\lambda^{-1}(\partial_x\mathsf r)
\\
\nonumber
&=\mathsf r^{-1}\mathsf r'
\\
\nonumber
&=\blockmatrix{\mathbf R^T\mathbf R'}{\mathbf R^T\mathbf r'}{\mathbf 0}{0}
\\
&=\blockmatrix{\mathbf U}{\mathbf u}{\mathbf 0}{0}.
\end{align}

Since $\mathds w(x,t)$ is a tangent vector at $\mathsf e$ for all $x$ and $t$, it makes sense to compute its derivative with respect to $x$ holding $t$ fixed. In order to compute this derivative, we observe that
\begin{align}
\nonumber
  0&=\left.\frac{\rm d}{{\rm d}\xi}\right|_{\xi=0}\mathsf r^{-1}(x+\xi,t)\mathsf r(x+\xi,t)
\\
\nonumber
&=\left.\frac{\rm d}{{\rm d}\xi}\right|_{\xi=0}\mathsf r^{-1}(x+\xi,t)\mathsf r(x,t)+
\left.\frac{\rm d}{{\rm d}\xi}\right|_{\xi=0}\mathsf r^{-1}(x,t)\mathsf r(x+\xi,t)
\\
&=\partial_x(\mathsf r^{-1})\mathsf r+\mathsf r^{-1}\partial_x\mathsf r,
\end{align}
whence
\begin{align}
  \partial_x(\mathsf r^{-1})=-\mathsf r^{-1}(\partial_x\mathsf r)\mathsf r^{-1}.
\end{align}
Consequently
\begin{align}
  \partial_x{\mathds w}&=\partial_x(\dot{\mathsf r}\mathsf r^{-1})=(\partial_x\dot{\mathsf r})\mathsf r^{-1}+\dot{\mathsf r}(\partial_x\mathsf r^{-1})\nonumber
\\
 &=(\partial_x\dot{\mathsf r})\mathsf r^{-1}-\dot{\mathsf r}\mathsf r^{-1}(\partial_x\mathsf r)\mathsf r^{-1}\nonumber\\
 &=\mathsf r(\mathsf r^{-1}\partial_x\dot{\mathsf r}-\mathsf r^{-1}\dot{\mathsf r}\mathsf r^{-1}(\partial_x\mathsf r))\mathsf r^{-1}\nonumber\\
 &=\mathsf r(\mathsf r^{-1}\partial_x\dot{\mathsf r}+\dot{\overline{\mathsf r^{-1}}}(\partial_x\mathsf r))\mathsf r^{-1}\nonumber
\\
\nonumber
&=\mathsf r \dot{\mathds u} \mathsf r^{-1}
\\
&=:{\rm Ad}_{\mathsf r}(\dot{\mathds u}).\label{eq:91}
\end{align}
Note that ${\rm Ad}_{\mathsf r}$ is the linear map induced on $T_{\mathsf e}\mathcal G$ by the adjoining operation ${\rm ad}_{\mathsf r}(\mathsf g)=\mathsf r\mathsf g\mathsf r^{-1}$. In matrix form, we have
\begin{align}
  {\rm Ad}_{\mathsf r}(\mathds u)&=\blockmatrix{\mathbf R}{\mathbf r}{\mathbf 0}{1}\blockmatrix{\mathbf U}{\mathbf u}{\mathbf 0}{0}\blockmatrix{\mathbf R^T}{-\mathbf R^T\mathbf r}{\mathbf 0}{1}\nonumber
\\
&=\blockmatrix{\mathbf R\mathbf U\mathbf R^T}{\mathbf R\mathbf u-\mathbf R\mathbf U\mathbf R^T\mathbf r}{\mathbf 0}{0}
\end{align}
It follows from \eqref{eq:6} that the differential operator $(\cdot)'$ does not commute with time derivative:
\begin{equation}
   \label{eq:20}
   \dot{\overline{\varphi'}}=\dot\varphi'-\mathring\lambda\varphi'.
 \end{equation}
On using \eqref{eq:91}, \eqref{eq:26}, and \eqref{eq:20}, we obtain:
\begin{align}
  {\mathds w}'&=(\dot{\mathsf r}\mathsf r^{-1})'=(\dot{\mathsf r})'\mathsf r^{-1}+\dot{\mathsf r}(\mathsf r^{-1})'\nonumber
\\
 &=(\dot{\mathsf r})'\mathsf r^{-1}-\dot{\mathsf r}\mathsf r^{-1}\mathsf r'\mathsf r^{-1}\nonumber\\
 &=\mathsf r(\mathsf r^{-1}(\dot{\mathsf r})'-\mathsf r^{-1}\dot{\mathsf r}\mathsf r^{-1}\mathsf r')\mathsf r^{-1}\nonumber\\
 &=\mathsf r(\mathsf r^{-1}(\dot{\overline{{\mathsf r}'}}-\mathring\lambda\mathsf r')-\mathsf r^{-1}\dot{\mathsf r}\mathsf r^{-1}\mathsf r')\mathsf r^{-1}\nonumber\\
 &=\mathsf r(\mathsf r^{-1}\dot{\overline{{\mathsf r}'}}+\dot{\overline{\mathsf r^{-1}}}\mathsf r')\mathsf r^{-1}-\mathring\lambda\mathsf r'\mathsf r^{-1}\nonumber
\\
\label{eq:95}
&={\rm Ad}_{\mathsf r}(\dot{\mathds u})-\mathring\lambda\mathds v,
\end{align}
an identity that is fundamental in our application of the principle of virtual powers. Indeed, on comparing \eqref{eq:95} and \eqref{eq:91}, we spot an extra term $\mathring\lambda\mathsf v$. It is exactly because of this term that the balance equation governing accretion contains contributions from the standard internal forces.

\subsection{Balance equations from the principle of virtual powers}
The power expended by the applied loads on a part $(a,b)\subset (0,1)$ has the representation: $\int_a^b \mathds b\cdot\mathds w{\rm ds}$, where
\begin{align}
  \mathds b=\blockmatrix{\mathbf B}{\mathbf b}{\mathbf 0}{0}
\end{align}
with $\mathbf B$ a skew-symmetric tensor and $\mathbf b$ a vector. The axial vector of $\mathbf B(x,t)$ represents the external moment per unit content applied at $x$ at time $t$, resolved with respect to the fixed pole $p_0$. For this, reason, we can refer to $\mathbf B$ as a couple. The vector $\mathbf b(x,t)$ is the external force per unit content applied at $x$ at time $t$. 

In addition to the external couple $\mathbf B$ and the external force $\mathbf b$, the part $(a,b)$ experiences an internal couple and an internal force transmitted by the rest of the body. We assume that these mechanical actions are conveyed by a pair of fields $\mathds t^+=\blockmatrix{\mathbf T^+}{\mathbf t^+}{\mathbf 0}0$ and $\mathds t^-=\blockmatrix{\mathbf T^-}{\mathbf t^-}{\mathbf 0}0$. The values of $\mathbf T^+$ and $\mathbf t^+$ at $b$ yield, respectively, the internal couple and the internal force exerted by $(b,1)$ on $(a,b)$. Likewise, the values of $\mathbf T^-$ and $\mathbf t^-$ at $a$ yield, respectively, the internal couple and the internal force exerted by $(0,a)$ on $(a,b)$. We therefore write the external power expended on $(a,b)$ as: 
\begin{align}\label{eq:53}
  \mathcal W_{\rm ext}[(a,b)]=\int_a^b \big(\mathds b\cdot\mathds w+{{\mathbf B_{\rm d}}}\cdot\dotUU+B\mathring\lambda\big)\,{\rm d}s+\mathds t^+(b)\cdot\mathds w(b)+\mathds t^-(a)\cdot\mathds w(a)
\end{align}
Here the dot product between skew tensors is defined by
\begin{align}
  \mathbf A\cdot\mathbf B=\frac 12 \mathbf A:\mathbf B,
\end{align}
where $:$ is the standard scalar product between second-order tensors. We recall the expression of the internal power:
\begin{equation}
  \label{eq:30}
\mathcal W_{\rm int}([a,b])=\int_a^b \big(\mathbf S\cdot\dot{\mathbf U}+\CC\cdot\dotUU+S\mathring\lambda\big){\rm d}s.
\end{equation}
We begin by noticing that, because of the inextensibilty constraint, we have $\dot{\mathbf u}=0$. Thus, the internal power may be written as:
\begin{equation}\label{eq:104}
\mathcal W_{\rm int}([a,b])=\int_a^b \big({\bm{\mathsf s}}\cdot\dot{\bm{\mathsf u}}+\CC\cdot\dotUU+S\mathring\lambda\big){\rm d}s.
\end{equation}
We now define
\begin{equation}
  \label{eq:98}
  {\bm{\mathsf t}}={\rm Ad}^*_{\mathsf r^{-1}}(\bm{\mathsf s}),
\end{equation}
where ${\rm Ad}^*_{\mathbf r}$ is the unique linear map such that
\begin{align}
  \mathds a\cdot{\rm Ad}_{\mathsf r}(\mathds b)={\rm Ad}_{\mathsf r}^*(\mathds a)\cdot\mathds b
\end{align}
for all $\mathds a=\blockmatrix{\mathbf A}{\mathbf a}{\mathbf 0}{0}$, $\mathds b=\blockmatrix{\mathbf B}{\mathbf b}{\mathbf 0}{0}$. Notice that \eqref{eq:98} entails
\begin{align}
  \mathds s=\blockmatrix{\mathbf S}{\mathbf s}{\mathbf 0}{0}&=\blockmatrix{\mathbf R^T(\mathbf T-\mathbf r\wedge\mathbf t)\mathbf R}{\mathbf R^T\mathbf t}{\mathbf 0}{0}={\rm Ad}^*_{\mathsf r}(\mathds t).
\end{align}
Now, we can write the internal power as:
\begin{align}\label{eq:120}
  \mathcal W_{\rm int}[(a,b)]=\int_a^b \big(\mathds t\cdot{\rm Ad}_{\mathsf r}(\dot{\mathds u})+\CC\cdot\dotUU+S\mathring\lambda\big)\,{\rm d}s.
\end{align}
Then, by \eqref{eq:95},
\begin{align}\label{eq:112}
  \mathcal W_{\rm int}[(a,b)]=\int_a^b \big(\mathds t\cdot\bm{\mathsf w}'+\CC\cdot\dotUU+(S-\mathds t\cdot\mathds v)\mathring\lambda\big)\,{\rm d}s.
\end{align}
On integrating by parts, we arrive at:
\begin{align}\label{eq:94}
  \mathcal W_{\rm int}[(a,b)]=\int_a^b \big(-\mathds t'\cdot\bm{\mathsf w}+\CC\cdot\dotUU+(S-\mathds t\cdot\mathds v)\mathring\lambda\big)\,{\rm d}s+\mathds t\cdot\mathds w|^b_{a}.
\end{align}
By imposing that internal and external powers be balanced for every virtual velocity, we obtain 
\begin{equation}\label{eq:122}
\begin{aligned}
&\int_a^b \big((\mathds t'+\mathds b)\cdot\bm{\mathsf w}+(\mathbf B_{\rm d}-\CC)\cdot\dotUU+(B-S+\mathds t\cdot\mathds v)\mathring\lambda\big)\,{\rm d}s\\
&\qquad\qquad\qquad +(\mathds t(b)-\mathds t^+(b))\cdot\mathds w(b)-(\mathds t(a)+\mathds t^-(a))\mathds w(a)=0.
\end{aligned}
\end{equation}
From the arbitrariness of $\mathds w$, $\dot{\mathbf U}_{\rm d}$, and $\mathring\lambda$, we obtain \eqref{eq:13}, and
\begin{align}\label{eq:93}
  \mathds t'+\mathds b=\mathds 0,
\end{align}
which is equivalent to \eqref{eq:15}. From the arbitrariness of parts, we also have the identification:
\begin{align}\label{eq:83}
   \mathds t=\mathds t^+=-\mathds t^-.
\end{align}

\section{Acknowledgements}
We are indebted with two anonymous reviewers, whose remarks and objections helped us to improve our paper. This research was supported by the Italian Agency INdAM--GNFM through the initiative: ``Progetto Giovani''.

%\bibliographystyle{abbrv}
%\bibliography{bibliography}
\end{document}